\documentclass[prb,aps,floats,twocolumn,superscriptaddress,floatfix]{revtex4}
\usepackage{graphicx,amssymb,amsmath,ifthen}

\begin{document}
%%%%%%%%%%%%%%%%%%%%%%%%%%%%%%%%%%%%%%%%%%%%%%%
\title{Quasiparticles governing the zero-temperature dynamics of the 1D
  spin-1/2 Heisenberg antiferromagnet in a magnetic field}
\author{Michael Karbach}
\affiliation{
  Bergische Universit{\"a}t Wuppertal, 
  Fachbereich Physik,
  D-42097 Wuppertal, Germany}
\author{Daniel Biegel}
\affiliation{
  Bergische Universit{\"a}t Wuppertal, 
  Fachbereich Physik,
  D-42097 Wuppertal, Germany}
\author{Gerhard M{\"u}ller}
\affiliation{
  Department of Physics, 
  University of Rhode Island, 
  Kingston RI 02881-0817}

%\ifthenelse{\equal{\writer}{gerhard}}%
%{\date{\today~--~2.0}} % Gerhard
%{\date{\version}}% Michael
\date{\today} 
%%%%%%%%%%%%%%%%%%%%%%%%%%%%%%%%%%%%%%%%%%%%%%%
\begin{abstract}
  The $T=0$ dynamical properties of the one-dimensional (1D) $s=\frac{1}{2}$
  Heisenberg antiferromagnet in a uniform magnetic field are studied via Bethe
  ansatz for cyclic chains of $N$ sites. The ground state at magnetization
  $0<M_z<N/2$, which can be interpreted as a state with $2M_z$ spinons or as a
  state of $M_z$ magnons, is reconfigured here as the vacuum for a different
  species of quasiparticles, the {\em psinons} and {\em antipsinons}. We
  investigate three kinds of quantum fluctuations, namely the spin fluctuations
  parallel and perpendicular to the direction of the applied magnetic field and
  the dimer fluctuations. The dynamically dominant excitation spectra are found
  to be sets of collective excitations composed of two quasiparticles excited
  from the psinon vacuum in different configurations.  The Bethe ansatz provides
  a framework for (i) the characterization of the new quasiparticles in relation
  to the more familiar spinons and magnons, (ii) the calculation of spectral
  boundaries and densities of states for each continuum, (iii) the calculation
  of transition rates between the ground state and the dynamically dominant
  collective excitations, (iv) the prediction of lineshapes for dynamic
  structure factors relevant for experiments performed on a variety of quasi-1D
  antiferromagnetic compounds, including KCuF$_3$,
  Cu(C$_4$H$_4$N$_2$)(NO$_3$)$_2$, and CuGeO$_3$.
\end{abstract}
%\pacs{??}
\maketitle
%%%%%%%%%%%%%%%%%%%%%%%%%%%%%%%%%%%%%%%%%%%%%%%
%
\section{Introduction}\label{sec:I}
%
%%%%%%%%%%%%%%%%%%%%%%%%%%%%%%%%%%%%%%%%%%%%%%%
Quantum spin chains are some of the most intensively studied models representing
strongly fluctuating quantum many-body systems because of their amenability to
exact analysis and because of the sustained interest in materials exhibiting
quasi-one-dimensional magnetic properties. Of particular interest are the
dynamical properties in the low-temperature regime, reflecting strong quantum
fluctuations.

Quantum fluctuations result from the time evolution of nonstationary observables
of a many-body system in the ground state.  They can be investigated
(experimentally, theoretically, or computationally) by dynamical probes. The
three main ingredients of each dynamical probe, (i) interaction Hamiltonian,
(ii) ground state, (iii) dynamical variable, make a specific set of collective
excitations visible to the probe. The specificity is determined by the
symmetries of all three ingredients.
  
A dynamical probe yields information on spectrum and transition rates. Different
sets of data are collected from the same many-body system [ingredients (i) and
(ii)] via particular fluctuation operators [ingredient (iii)]. Different
views of the quantum fluctuations are filtered out by operator specific
selection rules and transition rates.\cite{MK00}

Collective excitations are modes in which some of the tightly coupled
fundamental degrees of freedom (electrons, ions, atoms) move collectively in
more or less complex patterns. The free-particle like normal modes known to
exist in systems made of linearly coupled degrees of freedom are the inspiration
of attempts to describe collective excitations quite generally as composites of
elementary modes that are weakly coupled at most. This requires that the ground
state of the system can be meaningfully interpreted as a physical vacuum in
which certain kinds of elementary excitations (quasiparticles) move without
attenuation and scatter off each other nondestructively.

In completely integrable many-body systems, the identity of the quasiparticles in
any given eigenstate is upheld on a rigorous basis and encoded by a set of
quantum numbers. All excited states can then indeed be generated systematically
via the creation of quasiparticles from the ground state configured as a
physical vacuum. The interaction of the quasiparticles may not be weak, but it
is of a kind which preserves their identity. The factorizability of the
associated $S$-matrices, which is characteristic of completely integrable
systems, reduces all quasiparticle couplings to two-body scattering events for
which a general solution can be formulated, e.g. in the form of a Bethe wave
function.\cite{KBI93,GRS96}
  
The focus here is set on the quasiparticles which govern the quantum
fluctuations of the one-dimensional (1D) $s=\frac{1}{2}$ Heisenberg
antiferromagnet in an external magnetic field:\cite{Beth31,FT81}
\begin{equation}\label{eq:Hh}
H = \sum_{n=1}^N \left[J{\bf S}_n \cdot {\bf S}_{n+1}  - hS_n^z\right].
\end{equation}
The ground state at $h\geq h_S=2J$, $|F\rangle\equiv|\uparrow\uparrow\cdots\uparrow\rangle$, has saturated
magnetization, $M_z=N/2$. It is the reference state of the coordinate Bethe
ansatz and plays the role of the vacuum for {\em magnons} (spin-1
quasiparticles). All eigenstates of $H$ are described as configurations of
interacting magnons.  The ground state at $h=0$, $|A\rangle$, has magnetization
$M_z=0$. It contains $N/2$ magnons. In the framework of the Bethe ansatz, it is
reconfigured as the physical vacuum for {\em spinons}, a species of
spin-$\frac{1}{2}$ quasiparticles, and the entire spectrum of $H$ is
reinterpreted as composites of interacting spinon pairs. Likewise at
intermediate fields, $0<h<h_S$, the ground state $|G\rangle$ is reconfigured as a new
physical vacuum, and the low-lying collective excitations are most effectively
described as composites of two new species of quasiparticles, named {\em psinon}
and {\em antipsinon}.

In a recent paper,\cite{KM00} a detailed description of these quasiparticles in
the framework of the coordinate Bethe ansatz was given. Their role in the
zero-temperature spin fluctuations parallel to the direction of the magnetic
field was elucidated in the form of lineshape predictions for the associated
dynamic structure factor. Here we present a more comprehensive set of
applications, which also includes the perpendicular spin fluctuations and the
dimer fluctuations.

Physical realizations of Heisenberg antiferromagnetic chains have been known for
many years in the form of 3D crystalline compounds with quasi-1D exchange
coupling between magnetic ions. For the study of magnetic-field effects in the
dynamics as predicted in this paper, the coupling must not be too weak or else
it will be hard to reach the low-temperature regime. It must not be too strong
either or else it will be hard to reach a magnetic field that makes the Zeeman
energy comparable to the exchange energy.  One compound that promises to be
particularly suitable for this purpose is {\it copper pyrazine dinitrate}
[Cu(C$_4$H$_4$N$_2$)(NO$_3$)$_2$].\cite{HSR+99}

The spin fluctuations can be observed directly via inelastic neutron scattering
experiments. At very low temperatures, the dominant transitions in the
scattering experiment are between the ground state $|G\rangle$ and a set of
excitations $|\lambda\rangle$ that are reachable by one of the spin fluctuation operators
$S_q^\mu = N^{-1/2}\sum_n\,e^{iqn}S_n^\mu$, $\mu=x,y,z$. In the $T=0$ dynamic spin
structure factors
\begin{equation}\label{eq:dssf}
S_{\mu\mu}(q,\omega) = 2\pi\sum_\lambda|\langle G|S_q^\mu|\lambda\rangle|^2\delta\left(\omega-\omega_\lambda\right), 
\end{equation}
each transition with $\omega_\lambda\equiv E_\lambda-E_G$ and $q\equiv k_\lambda-k_G$
contributes a spectral line of intensity $2\pi|\langle G|S_q^\mu|\lambda\rangle|^2$.

Some quasi-1D antiferromagnetic compounds, of which CuGeO$_3$ is the most
prominent example,\cite{AFM+96,FKL+98,FL98} are susceptible to a spin-Peierls
transition, which involves a lattice distortion accompanied by an exchange
dimerization. The dimer fluctuations, $D_q = N^{-1/2}\sum_n\,e^{iqn}{\bf S}_n \cdot
{\bf S}_{n+1}$, as captured by the dynamic dimer structure factor
\begin{equation}\label{eq:ddsf}
S_{DD}(q,\omega) = 2\pi\sum_\lambda|\langle G|D_q|\lambda\rangle|^2\delta\left(\omega-\omega_\lambda\right)
\end{equation}
may not be as directly observable as the spin fluctuations but an understanding
of their quasiparticle composition is a matter of no less importance.

%%%%%%%%%%%%%%%%%%%%%%%%%%%%%%%%%%%%%%%%%%%%%%%
%
\section{Magnons, spinons, psinons}
\label{sec:msp}
%
%%%%%%%%%%%%%%%%%%%%%%%%%%%%%%%%%%%%%%%%%%%%%%%
The coordinate Bethe ansatz provides a natural classification of the eigenstates
of \eqref{eq:Hh} in terms of interacting magnons. The structure of the Bethe
wave function, its determination via the solution of the Bethe ansatz equations,
and its use for the calculation of matrix elements are summarized in the
Appendix.

For our discussion here it turns out to be sufficient to consider $r$-magnon
scattering states of the set $K_r$. In the invariant Hilbert subspace of
magnetization $M_z=N/2-r$, the Bethe quantum numbers of this set comprise, for
$0\leq r\leq N/2$ and $0\leq m\leq N/2-r$, all configurations
\begin{equation}\label{eq:I2msp}
-\frac{r}{2} + \frac{1}{2} -m \leq I_1 < I_2 < \cdots < I_r \leq \frac{r}{2} - \frac{1}{2} + m.
\end{equation}
The Bethe ansatz suggests a threefold interpretation of the ground state $|G\rangle$
at $0\leq M_z\leq N/2$ with quantum numbers
\begin{equation}\label{eq:IG}
\{I_i\}_G = \left\{-\frac{N}{4}+\frac{M_z}{2}+\frac{1}{2},\ldots,
\frac{N}{4}-\frac{M_z}{2}-\frac{1}{2} \right\}.
\end{equation}
Depending on the reference state (pseudo-vacuum) used, it can be regarded as a
scattering state of $N/2-M_z$ magnons, a scattering state of $2M_z$ spinons, or
the physical vacuum of psinons.\cite{KM00}

The states in the set $K_r$ then all contain the same number of magnons or
spinons but different numbers of psinons. The integer quantum number $m$ selects
all states from $K_r$ that contain $m$ pairs of psinons.  The ground state
$|G\rangle$ at $M_z=N/2-r$ is the only state with $m=0$. The quasiparticle role of
the psinons in the 2-psinon $(m=1)$ and 4-psinon $(m=2)$ scattering states was
highlighted previously.\cite{KM00}

The excitations which are important in $S_{zz}(q,\omega)$ (parallel spin
fluctuations) at $M_z=N/2-r$ were found to consist of a small subset of $K_r$
which includes $2m$-psinon states over the entire range $m$. However, all
$2m$-psinon states with significant spectral weight were found to belong to
particular configurations of Bethe quantum numbers $I_i$ in which $2m-1$ psinons
behave like a single degree of freedom with properties akin to those attributed
to an antiparticle. The spectrum of $S_{zz}(q,\omega)$ was thus identified as arising
predominantly from psinon-antipsinon $(\psi\psi^*)$ excitations.\cite{KM00} Here our
goal is to identify and interpret the dynamically relevant excitations also for
$S_{xx}(q,\omega) = \frac{1}{4}[S_{+-}(q,\omega) + S_{-+}(q,\omega)]$ (perpendicular spin
fluctuations) and $S_{DD}(q,\omega)$ (dimer fluctuations), where we expect psinons
and antipsinons to occur in different combinations.

%%%%%%%%%%%%%%%%%%%%%%%%%%%%%%%%%%%%%%%%%%%%%%%
%
\section{Symmetries and consequences}
\label{sec:selrul}
%
%%%%%%%%%%%%%%%%%%%%%%%%%%%%%%%%%%%%%%%%%%%%%%%
Narrowing down the dynamically dominant sets of excitations and characterizing
them as specific quasiparticle configurations proceeds in three steps. First we
limit the set of relevant excitations by the application of selection rules
which are imposed by the symmetry properties of the Hamiltonian \eqref{eq:Hh}
and the fluctuation operators $S_q^z,S_q^\pm,D_q$ and which are valid for
arbitrary system sizes. Then we identify additional selection rules that are
valid only for $N\to\infty$. Finally, we identify from the states not yet excluded
those whose transition rates are predominant in $S_{zz}(q,\omega)$, $S_{- +}(q,\omega)$,
$S_{+ -}(q,\omega)$, and $S_{DD}(q,\omega)$. This last step, which here is carried out
empirically, may very well find its ultimate justification by further symmetries
related to complete integrability.\cite{note1}

%%%%%%%%%%%%%%%%%%%%%%%%%%%%%%%%%%%%%%%%%%%%%%%
%
\subsection{Selection rules for arbitrary $N$}\label{sec:micro}
\begin{figure}[htb]
\centerline{\includegraphics[width=8.0cm,angle=0]{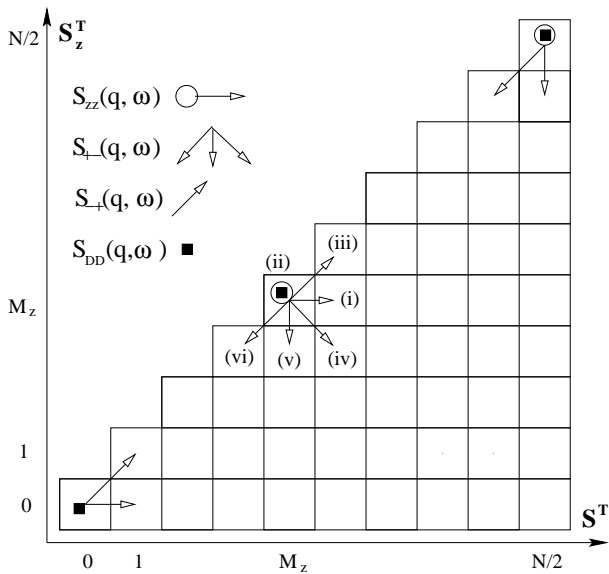}}

\caption{Transitions between the ground state $|G\rangle$ with quantum numbers
  $S_T=S_T^z=M_z$ and six classes of excitations permitted by microscopic
  selection rules (Ref.~\onlinecite{MTBB81}). Each class may contribute to
  exactly one of three dynamic spin structure factors. Class (ii) also
  contributes to the dynamic dimer structure factor. Fewer classes of
  permissible excitations exist for $M_z=0~ (h=0)$ and $M_z=N/2~ (h=h_S)$.}
\label{fig:excitation}
\end{figure}

%  
%%%%%%%%%%%%%%%%%%%%%%%%%%%%%%%%%%%%%%%%%%%%%%%
The conservation laws of the total spin $S_T$ and its $z$-component $S_T^z$
imply that transitions between eigenstates of \eqref{eq:Hh} induced by the
(nonstationary) spin fluctuation operators $S_q^z$, $S_q^\pm$ (vector) and the
dimer fluctuation operator $D_q$ (scalar) satisfy stringent selection rules.
The six classes of excitations with permissible transitions from $|G\rangle$ with
$S_T=S_T^z=M_z$ for the fluctuation operators $S_q^z,S_q^\pm,D_q$ are all listed
in Table~\ref{tab:classes}. The locations of these classes of excitations
relative to the ground state in the $(S_T,S_T^z)$-plane are shown in
Fig.~\ref{fig:excitation}.

%%%%%%%%%%%%%%%%%%%%%%%%%%%%%%BEGIN-TABLE%%%%%%
%\begin{widetext}
\begin{table*}[tb]
 \caption{Specifications of type-$K_r$ or equivalent states from six
   classes. Each class contains states that contribute to a specific dynamic
   spin structure factor at $T=0$. Class (ii) also contributes to the dynamic
   dimer structure factor. All specifications are relative to a given ground
   state with $S_T^z = S_T = N/2 - R = M_z$, where $M_z$ is the magnetization in
   a field of a certain strength $h$. The last column identifies the three
   subsets of excitations which dominate the spin and dimer fluctuations for
   $N\to\infty$.}\label{tab:classes}   
%\squeezetable  
\begin{tabular}{l|l|l|l|l|l|l}\hline\hline
class & $S_T$ & $S_{\mu\mu}(q,\omega)$ & $r$ & Bethe quantum numbers & 
Bethe ansatz solutions & dynamically\\
& $S_T^z$  &  &  &  & & dominant sets \\ \hline
(i) & $M_z+1$ & $S_{zz}(q,\omega)$ & $R$ & $I_i^{(i)} = I_i^{(iii)} +
\frac{1}{2}$, $i=1,\ldots,R-1$ & $z_i^{(i)} = z_i^{(iii)}$,
$i=1,\ldots,R-1$ & \\ 
& $M_z$ & & & $I_R^{(i)} = \frac{1}{2}(N-R+1)$ & $z_R^{(i)}=\infty$ &
 \\ \hline
(ii) & $M_z$ & $S_{zz}(q,\omega)$ & $R$ & $I_i^{(ii)}, i=1,\ldots,R$ &
$z_i^{(ii)}, i=1,\ldots,R$ & $\psi\psi^*$~ (P2) \\
& $M_z$ &$S_{DD}(q,\omega)$ & & from (\ref{eq:I2msp}) with $r=R$ & from (\ref{eq:bae}) with $r=R$
& \\ \hline
(iii) & $M_z+1$ & $S_{-+}(q,\omega)$ & $R-1$ & $I_i^{(iii)}, i=1,\ldots,R-1$
 & $z_i^{(iii)}, i=1,\ldots,R-1$ & $\psi\psi$~ (P3) \\
& $M_z+1$ & & & from (\ref{eq:I2msp}) with $r=R-1$ & from (\ref{eq:bae}) with 
$r=R-1$ & \\ \hline
(iv) & $M_z+1$ & $S_{+-}(q,\omega)$ & $R+1$ & $I_i^{(iv)} = I_i^{(iii)} +1$,
$i=1,\ldots,R-1$ & $z_i^{(iv)} = z_i^{(iii)}$, $i=1,\ldots,R-1$ & \\
& $M_z-1$ & & & $I_R^{(iv)} = I_{R+1}^{(iv)} = \frac{1}{2}(N-R)$ & $z_R^{(iv)}
= z_{R+1}^{(iv)} = \infty$  & \\ \hline
(v) & $M_z$ & $S_{+-}(q,\omega)$ & $R+1$ & $I_i^{(v)} = I_i^{(ii)} +
\frac{1}{2}$, $i=1,\ldots,R$ & $z_i^{(v)} = z_i^{(ii)}$, $i=1,\ldots,R$ & \\
& $M_z-1$ & & & $I_{R+1}^{(v)} = \frac{1}{2}(N-R)$ & $z_{R+1}^{(v)} = \infty$ & 
 \\ \hline
(vi) & $M_z-1$ & $S_{+-}(q,\omega)$ & $R+1$ & $I_i^{(vi)}, i=1,\ldots,R+1$
 & $z_i^{(vi)}, i=1,\ldots,R+1$ & $\psi\psi^*$~ (P6) \\
& $M_z-1$ & & & from (\ref{eq:I2msp}) with $r=R+1$ & from (\ref{eq:bae}) with 
$r=R+1$ & \\
\hline\hline
\end{tabular}
\end{table*}
%\end{widetext}
%%%%%%%%%%%%%%%%%%%%%%%%%%%%%END-TABLE%%%%%%%%%

%%%%%%%%%%%%%%%%%%%%%%%%%%%%%%%BEGIN-FIGURE%%%%
%%%%%%%%%%%%%%%%%%%%%%%%%%%%%%%%END-FIGURE%%%%%

We note that classes (ii), (iii), and (vi) include the sets $K_r$ for
$r=R,R-1,R+1$, respectively, while the remaining three classes (i), (iv), and
(v) include sets of states that belong to the same $S_T$ multiplets as the sets
$K_R$ or $K_{R+1}$. Table~\ref{tab:classes} also lists the Bethe quantum numbers
$I_i, i=1,\ldots,r$ for the type-$K_r$ or equivalent states and describes how the
rapidities $z_i, i=1,\ldots,r$ for these states are obtained.

%%%%%%%%%%%%%%%%%%%%%%%%%%%%%%%%%%%%%%%%%%%%%%%
%
\subsection{Selection rules for $N\to\infty$}\label{sec:macro}
%  
%%%%%%%%%%%%%%%%%%%%%%%%%%%%%%%%%%%%%%%%%%%%%%%
Before we begin evaluating matrix elements from Bethe ansatz solutions in
production mode, we take note that the rotational symmetry of the Hamiltonian
(\ref{eq:Hh}) and the vector nature of the spin fluctuation operator
$(S_q^x,S_q^y,S_q^z)$ imply the following rigorous relations between transition
rates involving excitations that belong to the same
$S_T$-multiplet:\cite{MTBB81}
\begin{subequations}\label{eq:msr}
\begin{eqnarray}\label{eq:msra}
|\langle G|S_q^z|\lambda^{(i)}\rangle|^2 &=&
\frac{|\langle G|S_q^-|\lambda^{(iii)}\rangle|^2}{2(M_z+1)}, \\ \label{eq:msrb}
|\langle G|S_q^+|\lambda^{(iv)}\rangle|^2 &=&
\frac{|\langle G|S_q^-|\lambda^{(iii)}\rangle|^2}{(M_z+1)(2M_z+1)}, \\ \label{eq:msrc}
|\langle G|S_q^+|\lambda^{(v)}\rangle|^2 &=&
\frac{2|\langle G|S_q^z|\lambda^{(ii)}\rangle|^2}{M_z}.
\end{eqnarray}
\end{subequations}

The significance of the relations (\ref{eq:msr}) is not limited to their
usefulness in reducing computational work. The magnetization is an extensive
quantity, implying $M_z\propto N$ at $h\neq 0$. All transition rates for class (i),
(iv), (v) excitations are then suppressed by factors $N$ or $N^2$ relative to
the transition rates of class (ii), (iii) excitations. The consequence is that
in a macroscopic system at $h\neq 0$, the spectral weight of all class (i), (iv),
(v) excitations in the $T=0$ dynamic spin structure factors $S_{\mu\mu}(q,\omega)$ is
negligible.\cite{note2}

%%%%%%%%%%%%%%%%%%%%%%%%%%%%%%%%%%%%%%%%%%%%%%%
%
\subsection{Selection rules related to integrability}\label{sec:integ}
%  
%%%%%%%%%%%%%%%%%%%%%%%%%%%%%%%%%%%%%%%%%%%%%%%
We shall find empirically that in each one of the remaining classes (ii), (iii),
and (vi) there exists a two-parameter set of excitations which governs one of
the dynamic structure factors of interest here. We shall name these sets P2, P3,
and P6, respectively (see Table~\ref{tab:classes}).  Corresponding finite-$N$
spectral contributions of class (i), (iv), and (v) states can be inferred from
Eqs.~(\ref{eq:msr}). The associated sets P1, P4, and P5 have their position in
the $(q,\omega)$-plane shifted vertically relative to the sets P2, P3, and P6
because of the Zeeman splitting, and the spectral weight of the former is
suppressed by factors $N$ or $N^2$ as explained previously.

%%%%%%%%%%%%%%%%%%%%%%%%%%%%%%%%%%%%%%%%%%%%%%%
%
\section{ Dynamically dominant excitations}
\label{sec:dde}
%  
%%%%%%%%%%%%%%%%%%%%%%%%%%%%%%%%%%%%%%%%%%%%%%%
In our search for the dynamically most relevant excitations, we focus on the
case of magnetization $M_z/N=\frac{1}{4}$ (half the saturation value). We
explore the transition rates for the spin and dimer fluctuation operators
between the ground state $|G\rangle$ and the type-$K_r$ states in the classes (ii),
(iii), (vi). These excitations are found to contribute most of the spectral
weight to the dynamic spin and dimer structure factors.

%%%%%%%%%%%%%%%%%%%%%%%%%%%%%%%%%%%%%%%%%%%%%%%
%
\subsection{Perpendicular spin fluctuations (P3)}
\label{sec:SC3}
\begin{figure}[b]
  \centerline{
\includegraphics[width=8.0cm]{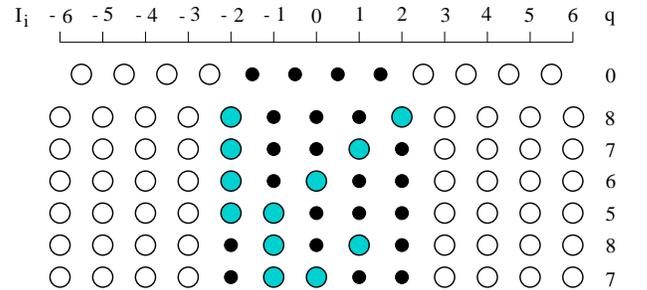}}

\caption{Psinon vacuum $|G\rangle$ for $N=16, M_z=4$ and $\psi\psi$ states
  with $q\geq0$ from the set $K_{3}$ out of class (iii). The $I_i$ values are
  marked by the positions of the magnons (small circles). The spinons (large
  circles) mark $I_i$-vacancies. A subset of the spinons are called psinons
  (gray circles) The wave numbers $q\equiv k-k_G$ are given in units of $2\pi/N$.}
\label{fig:pc3N16}
\end{figure}

%  
%%%%%%%%%%%%%%%%%%%%%%%%%%%%%%%%%%%%%%%%%%%%%%%
The spectral weight in $S_{- +}(q,\omega)$ is carried exclusively by class (iii)
excitations (see Table~\ref{tab:classes}). A systematic study of the transition
rates of type-$K_{R-1}$ states for $m=0,1,2,\ldots$ reveals that the dominant
contributions to the spectral weight come from two-psinon $(\psi\psi)$ states. The Bethe
quantum numbers of the states involved in these transitions are shown in
Fig.~\ref{fig:pc3N16} for $N=16$ and $M_z=4$. The top row represents the ground
state $|G\rangle$, which contains four magnons (small circles) or eight spinons
(large circles). This is the psinon vacuum at $R=4$.  The next row is the
lowest-lying two-psinon $(\psi\psi)$ state excited from $|G\rangle$. This excitation also
plays the role of the psinon vacuum $(m=0)$ at $R=3$. It is then characterized
as containing three magnons or ten spinons.

Mobilizing the two innermost spinons turns them into psinons (gray circles). The
remaining five rows in Fig.~\ref{fig:pc3N16} represent $\psi\psi$ states $(m=1)$ at
$0<q\leq\pi$ for $R=3$. The state in the second row is also counted as a $\psi\psi$
state. Hence there are six of them in total for $N=16$. Freeing up two
additional spinons from the sidelines produces a set of four-psinon states
$(m=2)$ of which the $\psi\psi$ states $(m=0,1)$ are special members. The maximum
number of psinons that can be mobilized at $R=3$ is equal to the number of
spinons: $2M_z=10$.

%%%%%%%%%%%%%%%%%%%%%%%%%%%%%%%BEGIN-FIGURE%%%%
%%%%%%%%%%%%%%%%%%%%%%%%%%%%%%%%END-FIGURE%%%%%

%%%%%%%%%%%%%%%%%%%%%%%%%%%%BEGIN-TABLE%%%%%%%%
\begin{table}[htb]
\caption{$\psi\psi$ states with $q\equiv k-k_G\geq0$ from the set $K_{R-1}$ out of class
  (iii) excited from the psinon 
  vacuum $|G\rangle$ for $N=24$, $R=6$: Bethe quantum numbers, wave 
  number (in units of $2\pi/N$), energy, and transition rate. The ground state has
  $k_G=0$ and $E_G=-11.5121346862$ and is realized at $h=1.58486\ldots$ for $N\to\infty$.}
  \begin{tabular}{rrrrrccc}\hline\hline
    & & $2I_i$ & & & $~ ~ ~q~ ~ ~$ & $~ ~E-E_G+h~ ~$ & $~ ~|\langle
    G|S_q^-|\lambda^{(iii)}\rangle|^2~ ~$ \\ \hline
    $-2$& +0& +2&+4&+6&      7&    1.8184194057&    0.0626472812\\
    $-4$& +0& +2&+4&+6&      8&    1.9556536638&    0.1424479327\\
    $-4$& $-2$& +2&+4&+6&      9&    2.0088932004&     0.2414561733\\
    $-4$& $-2$& +0&+4&+6&      10&    1.9746665911&    0.3895105465\\
    $-4$& $-2$& +0&+2&+6&      11&    1.8556790610&    0.6950450037\\
    $-4$& $-2$& +0&+2&+4&      12&    1.6606927007&    2.2311978057\\
    $-6$& +0& +2&+4&+6&      9&    2.1612403300&     0.0104636629\\
    $-6$& $-2$& +2&+4&+6&      10&    2.2113923024&    0.0348248694\\
    $-6$& $-2$& +0&+4&+6&      11&    2.1742818977&    0.0780182561\\
    $-6$& $-2$& +0&+2&+6&      12&    2.0527517142&    0.1743614335\\
    $-6$& $-4$& +2&+4&+6&      11&    2.3389550993&    0.0062315774\\
    $-6$& $-4$& +0&+4&+6&      12&    2.2987048423&    0.0260883689\\
    \hline\hline
  \end{tabular}
  \label{tab:pc3N24tr}
\end{table}
%%%%%%%%%%%%%%%%%%%%%%%%%%%%%%%%END-TABLE%%%%%%

\begin{figure}[b]

\centerline{
\includegraphics[width=6.5cm,angle=-90]{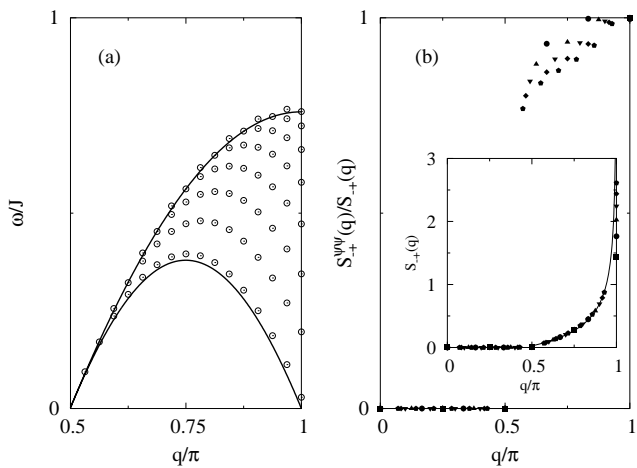}} 
\caption{(a) Energy versus wave number of the $\psi\psi$ excitations at $M_{z}=N/4$ for 
  $N=64$ (circles) and $N\to\infty$ (continuum P3 between curves). (b) Integrated
  intensity of $S_{- +}(q)$ (inset) and relative $\psi\psi$ contribution (main plot) for
  $N=12,16,20,24,28$.}
\label{fig:3}
\end{figure}

We have calculated the transition rates $|\langle G|S_q^-|\lambda^{(iii)}\rangle|^2$ between
the ground state for $N=24$, $M_z=4$ and all $2m$-psinon excitations. We found
that the $\psi\psi$ states are predominant. They are listed in
Table~\ref{tab:pc3N24tr} along with the momentum and energy transfer of the
associated spectral lines.  For $N\to\infty$ the $\psi\psi$ states form the continuum P3
in $(q,\omega)$-space.  In Fig.  \ref{fig:3}(a) we have plotted all states belonging
to P3 for $N=64$ (circles) and the spectral boundaries for $N\to\infty$. The range of
P3 is restricted to $\bar{q}_s\leq|q|\leq\pi$, where
\begin{equation}\label{eq:qsbar}
\bar{q}_s\equiv 2\pi M_z/N.
\end{equation}

Note that the continuum P3 is displaced by $\Delta q=\pi$ relative to the two-psinon
continuum discussed in the context of Ref.~\onlinecite{KM00}. The psinon vacuum
used for P3 is the state in the first row of Fig.~\ref{fig:pc3N16}. In
Ref.~\onlinecite{KM00} the state in the second row is the vacuum. These choices
are dictated by the different fluctuation operators considered now and then.

The relative integrated intensity of the $\psi\psi$ states, $S_{- +}^{\psi\psi}(q)/S_{-
  +}(q)$, is plotted in Fig.~\ref{fig:3}(b) for various $N$ at fixed $M_z=N/4$.
Corresponding data for the absolute integrated intensity $S_{- +}(q)$ are shown
in the inset.  We observe that there is virtually no intensity at $|q|\leq
\bar{q}_s$, outside the range of continuum P3. As $|q|$ increases from $q_s$
toward $\pi$, $S_{- +}(q)$ increases gradually and at an accelerated rate. The
value at the zone boundary diverges in the thermodynamic limit: $S_{- +}(\pi)\sim
N^{1-1/ \eta}$, with an exponent $\eta(M_z/N)$ that assumes the
value\cite{Hald80,FGM+96}
\begin{equation}\label{eq:eta}
  \eta(1/4)=1.53122\ldots
\end{equation}
for the situation at hand. It reflects the divergence $S_{- +}(q)\sim |\pi-q|^{1/
  \eta-1}$ for $N=\infty$. The relative $\psi\psi$ contribution to $S_{- +}(q)$ rises
rapidly from zero at $q\gtrsim q_s$ toward a value exceeding 97.8\% at $q=\pi$. The
solid line in the inset is obtained from a two-parameter fit, $a|\pi-q|^{1/
  \eta-1}+b$, of the data at $q\geq\pi/2$.

%%%%%%%%%%%%%%%%%%%%%%%%%%%%%%%BEGIN-FIGURE%%%%
%%%%%%%%%%%%%%%%%%%%%%%%%%%%%%%%END-FIGURE%%%%%

When we decrease $M_z$ at fixed $N$, the soft mode at $q=\pi$ remains stationary
while the soft mode at $q=\bar{q}_s$ moves to the left. At $M_z=0$, the $\psi\psi$
states become the two-spinon triplets. The two-spinon part of $S_{- +}(q,\omega)$ is
exactly known for $N\to\infty$.\cite{KMB+97,BKM98} Conversely, when we increase
$M_z$, the soft mode at $q=\bar{q}_s$ moves to the right and thus narrows the
range of P3 continually. At saturation $(M_z/N=\frac{1}{2})$ the function $S_{-
  +}(q,\omega)$ vanishes identically.

%%%%%%%%%%%%%%%%%%%%%%%%%%%%%%%%%%%%%%%%%%%%%%%
%
\subsection{Dimer and parallel spin fluctuations (P2)}
\label{sec:SC2}
\begin{figure}[b]
\centerline{
\includegraphics[width=7.5cm]{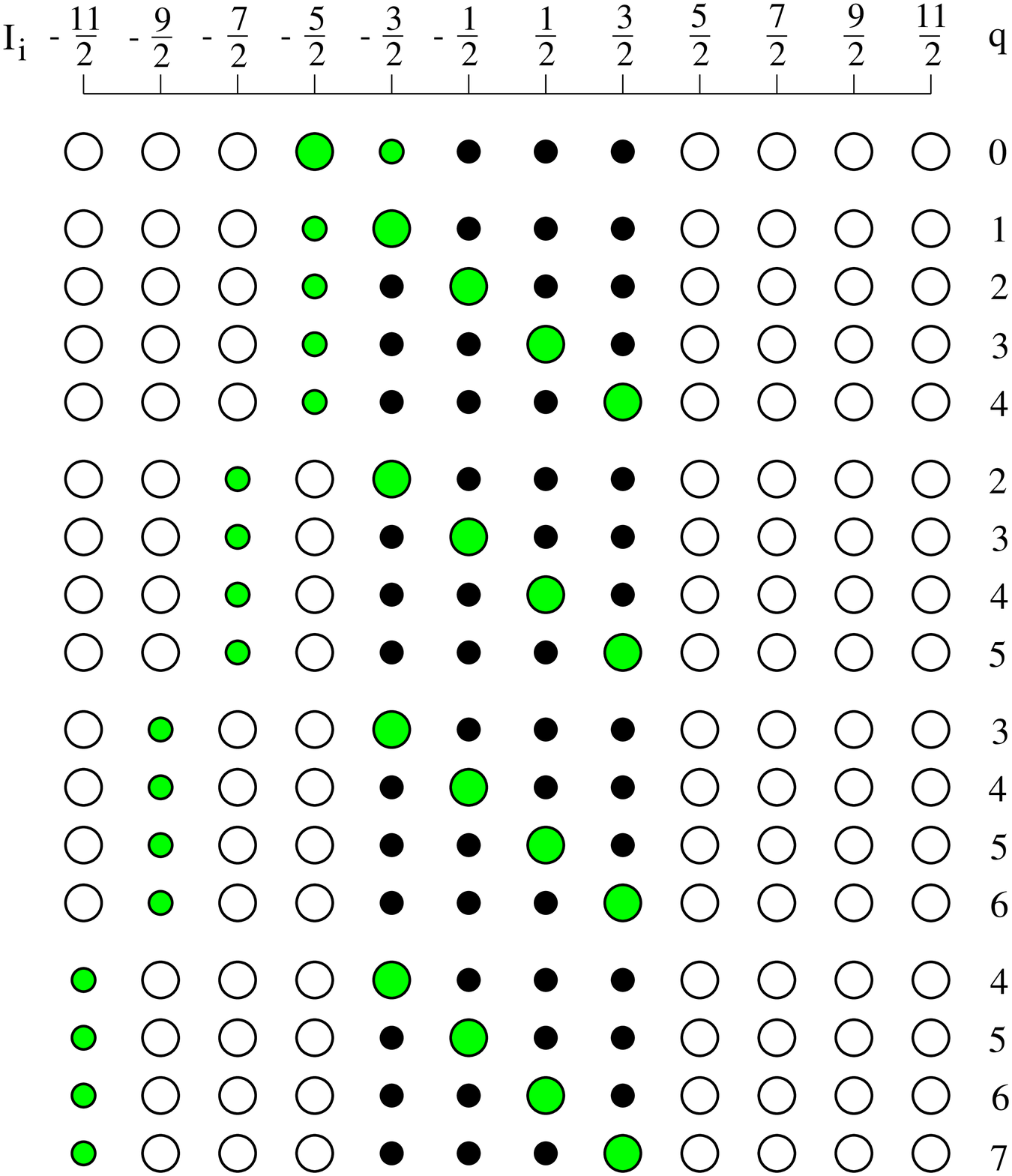}}
\caption{Psinon vacuum $|G\rangle$ for $N=16,M_z=4$ and set of $\psi\psi^*$ states with
  $0\leq q\leq \pi$ out of the set $K_4$.  The $I_i$ are given by the positions of
  the magnons (small circles) in each row. The spinons (large circles)
  correspond to $I_i$-vacancies. The psinon $(\psi)$ and the antipsinon $(\psi^*)$
  are marked by a large and a small grey circle, respectively. }
\label{fig:5}
\end{figure}

%  
%%%%%%%%%%%%%%%%%%%%%%%%%%%%%%%%%%%%%%%%%%%%%%%
The parallel spin fluctuations were already analyzed in Ref.~\onlinecite{KM00}.
The relevant excitations are contained in the set $K_R$ out of class (ii). This
set is subdivided into sets of $2m$-psinon states for $0\leq m\leq M_z=N/2-R$. Each
set for $m>0$ contributes one branch of excitations with significant spectral
weight to $S_{zz}(q,\omega)$.  Figure~\ref{fig:5} shows the configurations of Bethe
quantum numbers $I_i$ of all these states for $N=16, M_z=4$. The top row
represents the psinon vacuum $(m=0)$. The four groups of states underneath
represent the dynamically dominant branches of $2m$-psinon states for
$m=1,\ldots,4$.

%%%%%%%%%%%%%%%%%%%%%%%%%%%%%%%BEGIN-FIGURE%%%%
%%%%%%%%%%%%%%%%%%%%%%%%%%%%%%%%END-FIGURE%%%%%

\begin{figure}[bt]
\centerline{
\includegraphics[width=8.0cm]{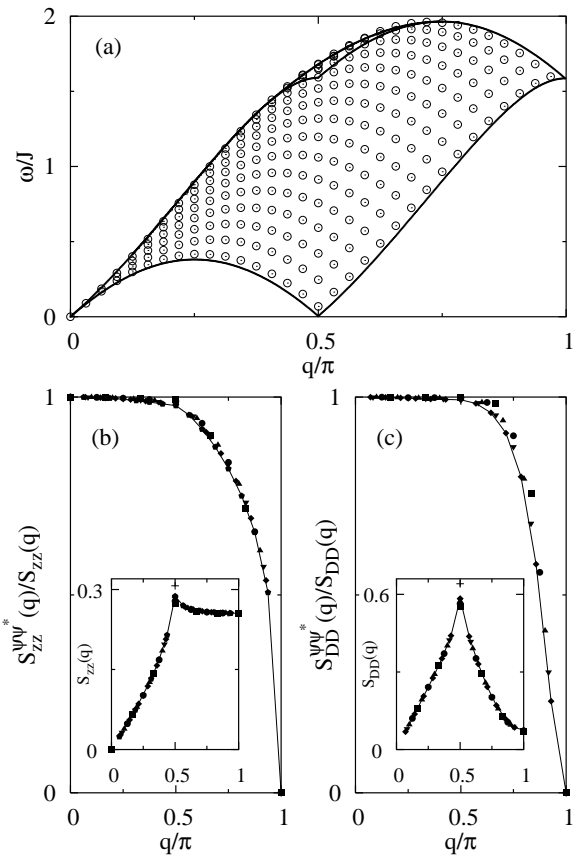}}
\caption{(a) Energy versus wave number the $\psi\psi^*$ excitations at $M_{z}/N=1/4$ for 
  $N=64$ (circles) and $N\to\infty$ (partially folded continuum P2 outlined by solid
  lines). (b) Integrated intensity $S_{zz}(q)$ (inset) and relative P2
  contribution (main plot) for $N=12, 16, 20, 24, 28, 32$. (c) Integrated
  intensity $S_{DD}(q)$ (inset) and relative P2 contribution (main plot) for
  $N=12, 16, 20, 24, 28$. The lines in (b) connect the $N=32$ data points and
  the lines in (c) the $N=28$ data points. The values of $S_{zz}(\pi/2)$ and
  $S_{DD}(\pi/2)$ extrapolated to $N\to\infty$ are marked (+).}
\label{fig:sc2}
\end{figure}

We argued that the $I_i$-configurations of these excitations suggest a simpler
interpretation in terms of two quasiparticles, namely one psinon $(\psi)$ and
one antipsinon $(\psi^*)$. We reinterpreted the series of dynamically dominant
branches taken from $2m$-parameter sets of multiple-psinon states as a single
two-parameter set of $\psi\psi^*$ scattering states.  These states form the
continuum P2 for $N\to\infty$ in the $(q,\omega)$-plane as illustrated in
Fig.~\ref{fig:sc2}(a). From the shape of the continuum with its soft modes at
$q=0$ and $q=q_s$, where 
\begin{equation}\label{eq:qs}
q_s\equiv\pi-2\pi M_z/N,
\end{equation}
and with the partial overlap along a stretch of the upper boundary, we
reconstructed the energy-momentum relations of the $\psi$ and $\psi^*$
quasiparticles.\cite{KM00}

%%%%%%%%%%%%%%%%%%%%%%%%%%%%%%%BEGIN-FIGURE%%%%
%%%%%%%%%%%%%%%%%%%%%%%%%%%%%%%%END-FIGURE%%%%%

%%%%%%%%%%%%%%%%%%%%%%%%%%%%BEGIN-TABLE%%%%%%%%
\begin{table}[tb]
\caption{$\psi\psi^*$ states with $q\equiv k-k_G\geq0$ from the set $K_{R}$ out of class
  (ii) excited from the psinon vacuum $|G\rangle$ for $N=16$, $R=4$: quantum number
  $m$, wave number (in units of $2\pi/N$), energy, and transition rates. The ground state has
  $k_G=0$ and $E_G=-11.5121346862$ and is realized at $h=1.58486\ldots$ for
  $N\to\infty$. The Bethe quantum numbers were listed in Table II of
  Ref.~\onlinecite{KM00}.}
\begin{center}
  \begin{tabular}{ccccc}\hline\hline
$2m$ & $~ ~q~ ~$ & $E-E_G$ & $~ ~|\langle G|S_q^z|\lambda^{(ii)}\rangle|^2~ ~$ & $~ ~|\langle
G|D_q|\lambda^{(ii)}\rangle|^2~ ~$\\ \hline
0 & 0 & 0.0000000000 & 1.0000000000 & 1.0000000000 \\
2 & 1 & 0.3504534152 & 0.0484825989 & 0.1201967890\\
2 & 2 & 0.5271937189 & 0.0587154211 & 0.1687346681\\
2 & 3 & 0.5002699273 & 0.0773592284 & 0.2298023543\\
2 & 4 & 0.2722787522 & 0.1257902349 & 0.3456324084\\
4 & 2 & 0.7981588810 & 0.0426892576 & 0.0720507048\\
4 & 3 & 0.9653287066 & 0.0552255878 & 0.1098585317\\
4 & 4 & 0.9301340415 & 0.0743667351 & 0.1555227849\\
4 & 5 & 0.6966798553 & 0.1253357676 & 0.2470269183\\
6 & 3 & 1.2708459328 & 0.0345439774 & 0.0307838904\\
6 & 4 & 1.4285177129 & 0.0516860817 & 0.0553527352\\
6 & 5 & 1.3858078992 & 0.0753564030 & 0.0866741700\\
6 & 6 & 1.1488426600 & 0.1406415212 & 0.1563073306\\
8 & 4 & 1.6819046570 & 0.0235815843 & 0.0060903835\\
8 & 5 & 1.8257803105 & 0.0443726010 & 0.0140423747\\
8 & 6 & 1.7724601200 & 0.0744641955 & 0.0259881320\\
8 & 7 & 1.5309413164 & 0.1686893882 & 0.0589091070\\
\hline\hline
  \end{tabular}
\end{center}
\label{tab:II}
\end{table}
%%%%%%%%%%%%%%%%%%%%%%%%%%%%%%%%END-TABLE%%%%%%

The corresponding search for the dynamically dominant dimer excitations again
points to the $\psi\psi^*$ continuum P2. The $\psi\psi^*$ transition rates for the
fluctuation operators $S_q^z$ and $D_q$ in a system with $N=16$, $M_z=4$ are
listed in Table~\ref{tab:II} for comparison. These data suggest that the
spectral weight in $S_{DD}(q,\omega)$ is concentrated more heavily at lower energy
than is observed in $S_{zz}(q,\omega)$. A more quantitative discussion of this
evidence will follow in Sec.~\ref{sec:V}.

Finite-$N$ data for the integrated intensities $S_{zz}(q)$ and $S_{DD}(q)$ are
presented in Figs.~\ref{fig:sc2}(b) and (c), respectively. Both static structure
factors rise from zero at $q=0$ to a cusp-like maximum at $q=q_s=\pi/2$, where
the soft mode is located. The cusp is of the form $\sim|q_s-q|^{\eta-1}$. A
two-parameter fit, $aN^{1-\eta}+b$, of the data at $q=q_s$ yields the extrapolated
values $S_{zz}(\pi/2)\simeq 0.307$ and $S_{DD}(\pi/2)\simeq0.641$. On approach to $q=\pi$,
the intensity drops more drastically in $S_{DD}(q)$ than in $S_{zz}(q)$.

For both kinds of fluctuations, the intensity at $q\leq q_s$ is almost exclusively
originating from $\psi\psi^*$ excitations. At $q=q_s$, the relative $\psi\psi^*$
contributions to $S_{zz}(q)$ and $S_{DD}(q)$ are estimated to be at least 93\%
and 95\%, respectively, in the limit $N\to\infty$.  At $q\gtrsim q_s$ the $\psi\psi^*$ parts
of $S_{zz}(q)$ and $S_{DD}(q)$ decrease monotonically but remain dominant except
in the immediate vicinity of the zone boundary. The data suggest a qualitative
difference in how the relative $\psi\psi^*$ intensities approach zero as $q\to\pi$. If
the behavior near the zone boundary can be described by a power-law,
$\sim|\pi-q|^\gamma$, then we predict $\gamma\gtrsim 1$ for the dimer fluctuations and $\gamma\simeq
0.3$ for the spin fluctuations.\cite{KM00}

Upon varying the value of $M_z$, the continuum P2 changes its shape
continuously. In both limits, $M_z/N\to0$ and $M_z/N\to\frac{1}{2}$, it
degenerates into a single branch and then vanishes.\cite{KM00,MTBB81} At $M_z=0$
the dimer fluctuations and the spin fluctuations are produced by entirely
different sets of excitations. $S_{zz}(q,\omega)$ is known to be dominated by the
continuum of two-spinon triplet excitations,\cite{KMB+97} which are Bethe ansatz
solutions with real rapidities. $S_{DD}(q,\omega)$ is presumably governed by
two-spinon singlet excitations, which are Bethe ansatz solutions with complex
rapidities.\cite{KHM98}

%%%%%%%%%%%%%%%%%%%%%%%%%%%%%%%%%%%%%%%%%%%%%%%
%
\subsection{Perpendicular spin fluctuations (P6)}\label{sec:SC6}
%  
%%%%%%%%%%%%%%%%%%%%%%%%%%%%%%%%%%%%%%%%%%%%%%%
For finite $N$, the spectral weight in the dynamic structure factor
$S_{+-}(q,\omega)$ probes excitations from classes (iv)-(vi). However, we know from
Sec.~\ref{sec:macro} that the intensities of class-(iv) and class-(v)
excitations are bound to fade away in the limit $N\to\infty$. The class-(vi)
excitations contain the set $K_{R+1}$. It turns out that much of the spectral
weight in $S_{+-}(q,\omega)$ is carried again by $\psi\psi^*$ excitations.  However, the
transitions $\langle G|S_q^+|\lambda\rangle$ probe the $\psi\psi^*$ states in a different invariant
subspace than the transitions $\langle G|S_q^z|\lambda\rangle$, $\langle G|D_q|\lambda\rangle$ do. This causes
some dramatic changes in the spectrum and in the spectral-weight distribution.

%%%%%%%%%%%%%%%%%%%%%%%%%%%%%%%BEGIN-FIGURE%%%%
\begin{figure}[b]
\includegraphics[width=8.0cm]{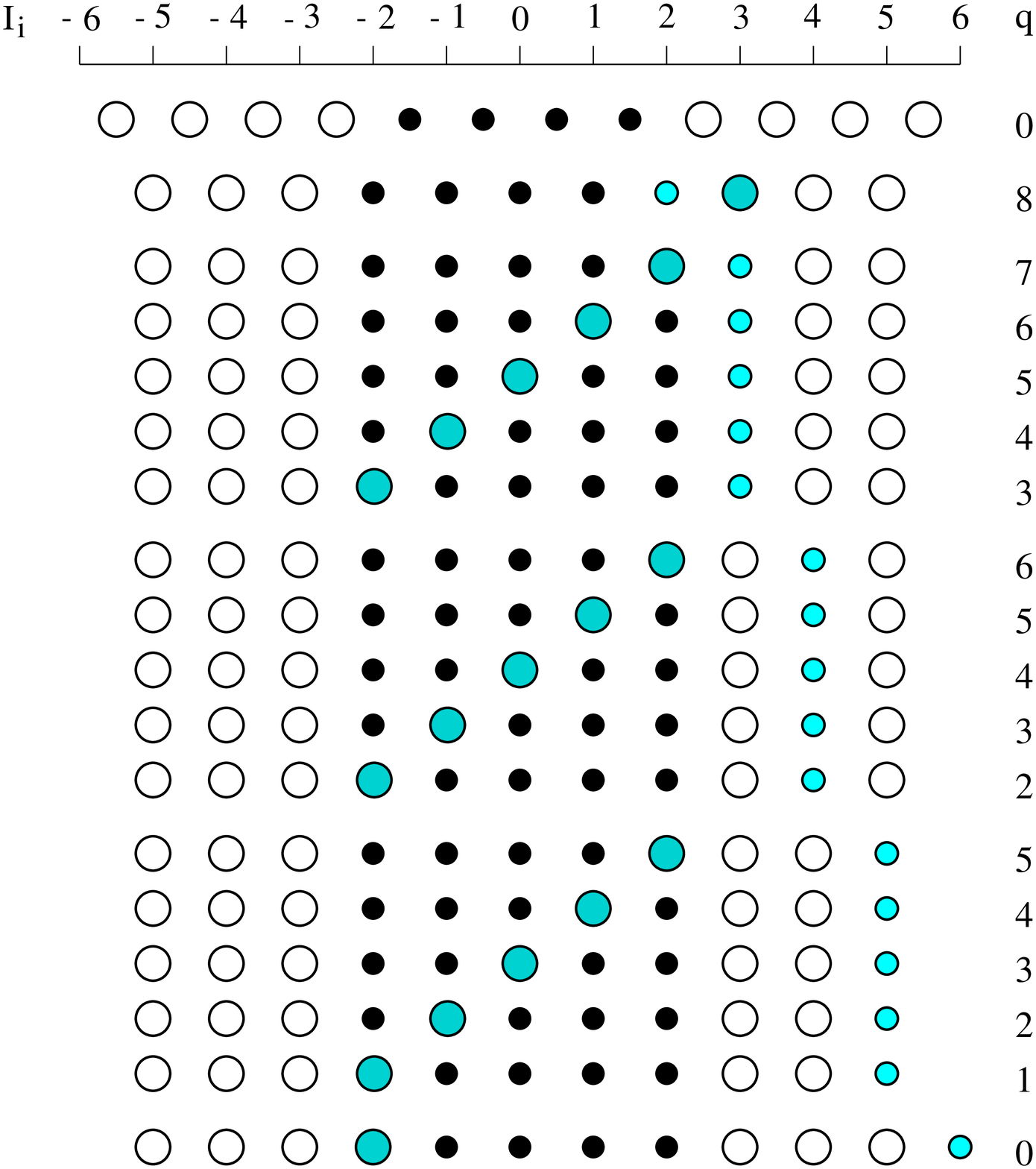}

\caption{Psinon vacuum $|G\rangle$ for $N=16,M_z=4$ and set of $\psi\psi^*$ states with
  $0\leq q\leq \pi$ out of the set $K_5$. The $I_i$ are given by the positions of
  the magnons (small circles) in each row. The spinons (large circles)
  correspond to $I_i$-vacancies. The psinon $(\psi)$ and the antipsinon $(\psi^*)$
  are marked by a large and a small grey circle, respectively. The last row
  describe a state from class (v) which belongs to the same $S_T$-multiplet as
  the psinon vacuum.}
\label{fig:pc6bqn}
\end{figure}
%%%%%%%%%%%%%%%%%%%%%%%%%%%%%%%%END-FIGURE%%%%%

The differences are best illustrated by Fig.~\ref{fig:pc6bqn} in relation to
Figs.~\ref{fig:pc3N16} and \ref{fig:5}. The top row in all three figures shows
the $I_i$-configuration of the ground state $|G\rangle$ for $N=16$, $M_z=4$. The
remaining rows in Fig.~\ref{fig:5} represent the $\psi\psi^*$ states in the same
invariant subspace, whereas the remaining rows in Fig.~\ref{fig:pc3N16}
represent the $\psi\psi$ states in the subspace with one less magnon (i.e. two
more spinons). The first $\psi\psi$ state (second row in Fig.~\ref{fig:pc3N16})
also plays the role of the psinon vacuum in that subspace.

\begin{figure}[tb]
\includegraphics[width=6.0cm,angle=-90]{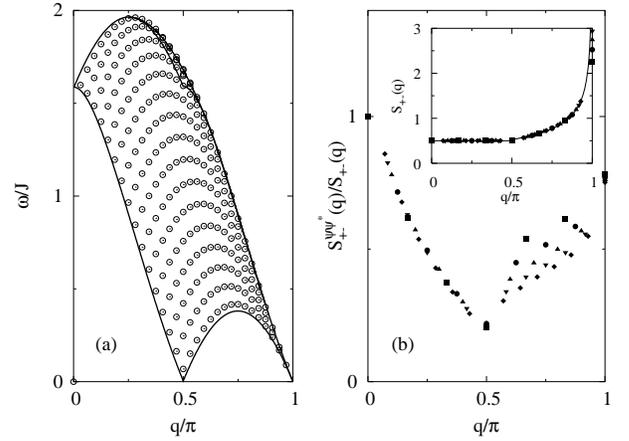}
\caption{(a) Energy versus wave number the $\psi\psi^*$ excitations at $M_{z}=N/4$ for 
  $N=64$ (circles) and $N\to\infty$ (partially folded continuum P6 outlined by solid
  lines). (b) Integrated intensity $S_{+ -}(q)$ (inset) and relative P6
  contribution (main plot) for $N=12, 16, 20, 24, 28$. The solid line in the
  inset results from a two-parameter fit as explained in the context of
  Fig.~\ref{fig:3}.}
\label{fig:pc6spec}
\end{figure}

The second row in Fig.~\ref{fig:pc6bqn} represents the lowest excitation probed
by $\langle G|S_q^+|\lambda\rangle$ and, at the same time, the psinon vacuum with one more magnon
(i.e. two less spinons). The three groups of five states underneath represent
the complete set of $\psi\psi^*$ states in the same invariant subspace. Because the
momentum transfer for these $\psi\psi^*$ states is relative to a different
psinon vacuum than was the case for the $\psi\psi^*$ states discussed in
Fig.~\ref{fig:5}, the observable spectrum of the continuum P6 which emerges for
$N\to\infty$ [Fig.~\ref{fig:pc6spec}(a)] is the mirror image of the continuum P2
[Fig.~\ref{fig:sc2}(a)]. 

%%%%%%%%%%%%%%%%%%%%%%%%%%%%%%%BEGIN-FIGURE%%%%
%%%%%%%%%%%%%%%%%%%%%%%%%%%%%%%%END-FIGURE%%%%%

%%%%%%%%%%%%%%%%%%%%%%%%%%%%BEGIN-TABLE%%%%%%%%
\begin{table}[tb]
\caption{$\psi\psi$ states with $q\equiv k-k_G\geq0$ from the set $K_{R+1}$ out of class
  (vi) excited from the psinon vacuum $|G\rangle$ for $N=16$, $R=4$: Bethe quantum
  numbers, quantum number $m$, wave number (in units of $2\pi/N$), energy, and
  transition rate. The ground state has $k_G=0$ and $E_G=-11.5121346862$ and is
  realized at $h=1.58486\ldots$ for $N\to\infty$. The last row describe a state from
  class (v) which belongs to the same $S_T$-multiplet as $|G\rangle$.}
\begin{center}
  \begin{tabular}{ccccc}\hline\hline
$2I_i$ & $2m$ & $q$ & $E-E_G-h$ & $|\langle G|S_q^+|\lambda^{(vi)}\rangle|^2$ \\ \hline
$-4-2+0+2+4$ & 0 & 8 & -1.4624484093 & 1.9420228564\\
$-4-2+0+2+6$ & 2 & 7 & -1.0239463125  & 0.6324984574      \\
$-4-2+0+4+6$ & 2 & 6 & -0.7427954774  & 0.0473348211 \\
$-4-2+2+4+6$ & 2 & 5 & -0.6661027722  & 0.0187604165  \\
$-4+0+2+4+6$ & 2 & 4 & -0.8076063179  & 0.0156977974 \\
$-2+0+2+4+6$ & 4 & 3 & -1.1468483618  & 0.0980201222 \\
$-4-2+0+2+8$ & 4 & 6 & -0.5201660070  & 0.3553105587 \\
$-4-2+0+4+8$ & 4 & 5 & -0.2494575408 & 0.0606512819 \\
$-4-2+2+4+8$ & 4 & 4 & -0.1822770670  & 0.0309111714 \\
$-4+0+2+4+8$ & 6 & 3 & -0.3302018783  & 0.0280474732 \\
$-2+0+2+4+8$ & 6 & 2 & -0.6710449481  & 0.2106074048     \\
$-4-2+0+2+10$ & 6 & 5 & -0.0400322092  & 0.1984683911 \\
$-4-2+0+4+10$ & 6 & 4 & +0.2167505272  & 0.0646015620 \\
$-4-2+2+4+10$ & 8 & 3 & +0.2717353551  & 0.0401817835 \\
$-4+0+2+4+10$ & 8 & 2 & +0.1166329921  & 0.0372990435 \\
$-2+0+2+4+10$ & 8 & 1 & -0.2237883038 & 0.3577163008 \\
$-4-2+0+2+12$ & 8 & 0 &  0.0000000000 & 0.5000000000 \\
\hline\hline
  \end{tabular}
\end{center}
\label{tab:pc6}
\end{table}
%%%%%%%%%%%%%%%%%%%%%%%%%%%%%%%%END-TABLE%%%%%%

The last row in Fig.~\ref{fig:pc6bqn} is not exactly a $\psi\psi^*$ state.  It
differs from the $\psi\psi^*$ in the previous row only by the smallest change in one
Bethe quantum number. This class-(v) state belongs to the same $S_T$-multiplet
as the ground state $|G\rangle$ (top row). We have included it here because its
transition rate is significant. In fact, it is the only excitation at $q=0$ with
a nonzero transition rate. Even though it is not a member of the set $K_{R+1}$,
its contribution to $S_{+-}(q,\omega)$ marks a natural endpoint of the continuum P6.
The excitation energies and transition rates pertaining to all states shown in
Fig.~\ref{fig:pc6bqn} are listed in Table~\ref{tab:pc6}.

The integrated intensity $S_{+ -}(q)$ as shown in Fig.~\ref{fig:pc6spec}(b) is
almost flat in the region $0\leq q\leq q_s$. The intensity at $q=0$ is exactly
known:\cite{MTBB81}
\begin{equation}\label{eq:spm0}
S_{+ -}(0) = 2M_z/N=1/2.
\end{equation}
At $q>\bar{q}_s$ the function $S_{+ -}(q)$ rises gradually and with increasing
slope ending in a divergence at $q=\pi$. The relation 
\begin{equation}\label{eq:spmmp}
S_{+ -}(q) = 2M_z/N +S_{- +}(q)
\end{equation}
dictates that the singularity is the one already described in
Sec.~\ref{sec:SC3}: $S_{+ -}(q)\sim |\pi-q|^{1/ \eta-1}$.

The relative $\psi\psi^*$ contribution to the integrated intensity is largest near
the zone center and near the zone boundary as shown in the inset to
Fig.~\ref{fig:pc6spec}(b). It gradually drops from 100\% at $q=0$ to $\sim20\%$ at
the soft-mode position $q=\bar{q_s}$ and then rises back to $\sim72.5\%$ at
$q=\pi$.  Note that the $N$-dependence of the relative intensity data is much
stronger at $q>\bar{q_s}$ than at $q\leq\bar{q_s}$. We shall see that the
qualitatively different $N$-dependences are also observed in transition rates,
from which interesting conclusions can be drawn.

%%%%%%%%%%%%%%%%%%%%%%%%%%%%%%%%%%%%%%%%%%%%%
%
\section{Lineshapes}\label{sec:V}
%
%%%%%%%%%%%%%%%%%%%%%%%%%%%%%%%%%%%%%%%%%%%%%

To calculate the lineshapes of the $\psi\psi$ and $\psi\psi^*$ contributions to the
dynamic spin and dimer structure factors we use, wherever applicable, the
product ansatz
\begin{equation}\label{eq:prorep}
S(q,\omega)=D(q,\omega)M(q,\omega)
\end{equation}
discussed at some length in Ref.~\onlinecite{KM00}. The factor $D(q,\omega)$ is the
density of $\psi\psi$ or $\psi\psi^*$ states, which can be evaluated for very large $N$
via Bethe ansatz. The factor $M(q,\omega)$ represents the scaled transition rates
$N|\langle G|S_q^\mu|\lambda\rangle|^2$, $\mu=z,+,-$, or $N|\langle G|D_q|\lambda\rangle|^2$ between the ground
state and the sets of $\psi\psi$ or $\psi\psi^*$ states. These matrix elements are also
calculated via Bethe ansatz but only for much smaller systems.

For the applications considered here, the product ansatz depends on a reasonably
fast convergence, within the spectral boundaries of the continua P2, P3, and P6,
of the finite-$N$ transition rate data toward a smooth function $M(q,\omega)$ as
$N\to\infty$. Problems with this ansatz arise when the finite-size excitations for
which transition rates are available are subject to significant energy shifts
caused by the quasiparticle interaction. For $\psi\psi$ and $\psi\psi^*$ scattering
states, these are effects of O$(N^{-1})$ as discussed in Ref.~\onlinecite{KM00}.
The exercise of caution is also indicated when the scaling behavior of the
finite-size transition rates changes at spectral boundaries as is frequently the
case.

Notwithstanding these caveats, the product ansatz is a useful tool for merging
the best available transition-rate data and density-of-states data. It was
successfully tested for the two-spinon excitations at $M_z=0$.\cite{KMB+97} Any
major distortions of the lineshapes predicted by the product ansatz can be
avoided if we omit all data points of $M(q,\omega)$ that are shifted across spectral
boundaries. Any theoretical and computational advances that make it possible to
calculate transition rates for larger systems will improve the predictive power
of the product ansatz.

%%%%%%%%%%%%%%%%%%%%%%%%%%%%%%%%%%%%%%%%%%%%%%%
%
\subsection{$S_{DD}(q,\omega)$ and $S_{zz}(q,\omega)$}
\label{sec:P2}
%  
%%%%%%%%%%%%%%%%%%%%%%%%%%%%%%%%%%%%%%%%%%%%%%%

\begin{figure}[tb]
  \centering
  \includegraphics[width=6.0cm,angle=-90]{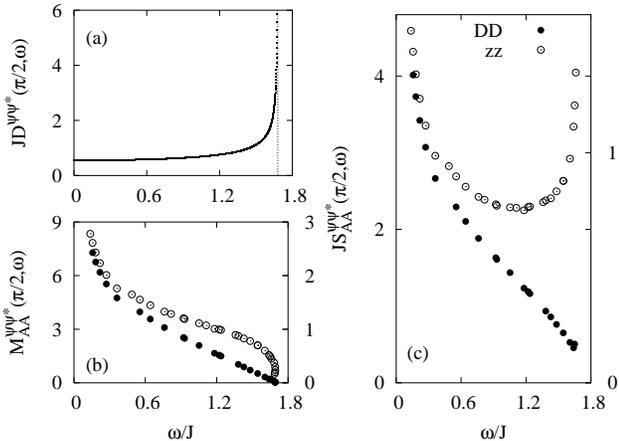}

\caption{(a) Density of $\psi\psi^*$ states at $q=\pi/2$ for $N=2048$. (b) Dimer
  ($DD$) and parallel spin ($zz$) transition rates between the psinon vacuum and
  the $\psi\psi^*$ states at $q=\pi/2$ for $N\leq32$ (zz) and $N\leq28$ (DD). (c)
  Lineshape at $q=\pi/2$ of the $\psi\psi^*$ contribution to $S_{zz}(q,\omega)$ and
  $S_{DD}(q,\omega)$. All results are for $M_z=N/4$. The scales in (b) and (c) on
  the left (right) are for $DD$ ($zz$).}
\label{fig:sddpd2}
\end{figure}

What can be observed in a fixed-$q$ scan of the dimer and parallel spin
fluctuations at $q=q_s$? The lineshape determination of
$S_{DD}^{\psi\psi^*}(\pi/2,\omega)$ is illustrated in Fig.~\ref{fig:sddpd2} in comparison
with corresponding data for $S_{zz}^{\psi\psi^*}(\pi/2,\omega)$ as shown in Fig.~8 of
Ref.~\onlinecite{KM00}.  The first factor in the product ansatz is the same for
both sets of data, namely the density of $\psi\psi^*$ states in continuum P2. It has
the characteristic shape with a square-root divergence at the upper band edge
$\omega_U$ as shown in panel (a).

%%%%%%%%%%%%%%%%%%%%%%%%%%%%%%%BEGIN-FIGURE%%%%
%%%%%%%%%%%%%%%%%%%%%%%%%%%%%%%%END-FIGURE%%%%%
  
The scaled transition rates $M_{zz}^{\psi\psi^*}(\pi/2,\omega)$ and
$M_{DD}^{\psi\psi^*}(\pi/2,\omega)$ [panel (b)] are monotonically decreasing functions.
For both kinds of fluctuations the data at low frequencies are consistent with
the power-law divergence,
\begin{equation}\label{eq:sddzzom0}
S_{zz}^{\psi\psi^*}(\pi/2,\omega) \sim S_{DD}^{\psi\psi^*}(\pi/2,\omega) \sim\omega^{\eta-2} 
\end{equation}
with $\eta-2=-0.468\ldots$ as predicted by conformal invariance.  It is near $\omega_U$
where the two sets of data differ most. While both transition rate functions
tend to vanish at $\omega_U$, this tendency is considerably slower for the parallel
spin fluctuations than for the dimer fluctuations. The resulting lineshapes are
dramatically different.

We saw that the slow approach to zero at $\omega_U$ of the parallel spin transition
rates $M_{zz}^{\psi\psi^*}(\pi/2,\omega)$ combined with the divergence in
$D^{\psi\psi^*}(\pi/2,\omega)$ produces a diverging trend at $\omega_U$ in
$S_{zz}^{\psi\psi^*}(\pi/2,\omega)$. The result is a characteristic double-peak
structure.\cite{KM00} The more rapid approach to zero of the dimer transition
rates $M_{DD}^{\psi\psi^*}(\pi/2,\omega)$ overcomes the divergence of
$D^{\psi\psi^*}(\pi/2,\omega)$ and produces a single-peak lineshape in
$S_{DD}^{\psi\psi^*}(\pi/2,\omega)$.

\begin{figure}[bt]
  \centerline{
\includegraphics[width=6.0cm,angle=-90]{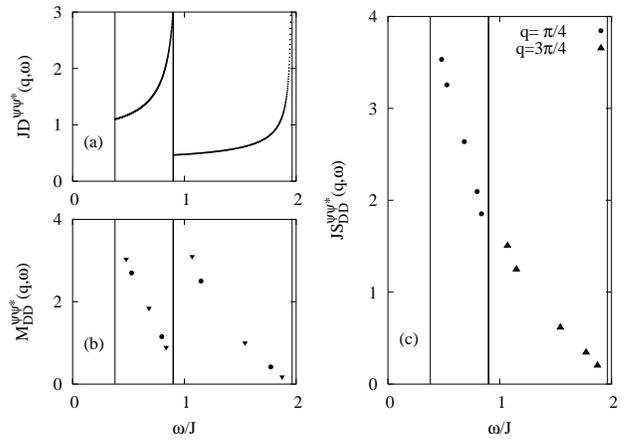}}

\caption{(a) Density of $\psi\psi^*$ states at $q=\pi/4,3\pi/4$ for $N=2048$. (b)
  Dimer transition rates between the psinon vacuum and the $\psi\psi^*$ states at
  $q=\pi/4,3\pi/4$ for $N=16,24$. (c) Lineshape at $q=\pi/4,3\pi/4$ of the $\psi\psi^*$
  contribution to $S_{DD}(q,\omega)$. All results are for $M_z=N/4$.}
\label{fig:sddqpid4}
\end{figure}

Now we consider the wave numbers halfway between the soft mode $q_s$ and the
zone boundary or the zone center. The lineshape determination via product ansatz
of $S_{DD}^{\psi\psi^*}(q,\omega)$ at $q=\pi/4$, $3\pi/4$ is illustrated in
Fig.~\ref{fig:sddqpid4}. The corresponding data for the parallel spin
fluctuations were shown in Fig.~9 of Ref.~\onlinecite{KM00}. For this situation
the upper edge of one band $(q=\pi/4)$ coincides with the lower edge of the other
band $(q=3\pi/4)$.

%%%%%%%%%%%%%%%%%%%%%%%%%%%%%%%BEGIN-FIGURE%%%%
%%%%%%%%%%%%%%%%%%%%%%%%%%%%%%%%END-FIGURE%%%%%

Even though the number of available data points for dimer transition rates is
limited, there is a clear indication that the lineshapes of the dimer
fluctuations again consist of single-peak structures with a divergence at the
lower band edge $\omega_L\neq 0$ and a shoulder reaching to the upper band edge
$\omega_U$, in strong contrast to the double-peak structures predicted for the
lineshapes of the parallel spin fluctuations.

In summary, the spectral-weight distribution of the dynamic structure factors
which probe the dimer and parallel spin fluctuations have many commonalities but
also some very distinct properties. In both cases, the dominant spectrum is the
continuum P2 of $\psi\psi^*$ states within the invariant $M_z$ subspace which also
contains the ground state $|G\rangle$. Both functions $S_{DD}^{\psi\psi^*}(q,\omega)$ and
$S_{zz}^{\psi\psi^*}(q,\omega)$ are strongly peaked along the lower continuum
boundary $\omega_L(q)$. Only the latter is also peaked along the upper continuum
boundary $\omega_U(q)$. The divergence along $\omega_L(q)$ is caused by the transition
rates, whereas the divergence along $\omega_U(q)$ is a density-of-states effect.

%%%%%%%%%%%%%%%%%%%%%%%%%%%%%%%%%%%%%%%%%%%%%%%
%
\subsection{$S_{- +}(q,\omega)$}
\label{sec:P3}
%  
%%%%%%%%%%%%%%%%%%%%%%%%%%%%%%%%%%%%%%%%%%%%%%%

\begin{figure}[b]
 \centerline{
\includegraphics[width=6.5cm,angle=-90]{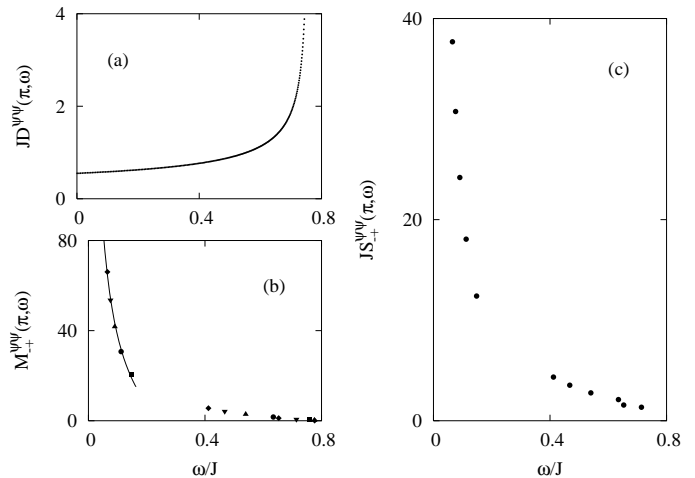}}

\caption{(a) Density of $\psi\psi$ states at $q=\pi$ for $N=2048$. (b) Perpendicular
  spin transition rates between the psinon vacuum and the $\psi\psi$ states at
  $q=\pi$ for $N=12, 16, 20, 24, 28$. (c) Lineshape at $q=\pi$ of the $\psi\psi$
  contribution to $S_{- +}(q,\omega)$. All results are for $M_z=N/4$.}
\label{fig:s-+pi}
\end{figure}

Here we focus on the lineshape at $q=\pi$ of the $\psi\psi$ contribution to the
dynamic spin structure factor $S_{- +}(q,\omega)$. The continuum P3 of $\psi\psi$ states
was previously found to be dominant. The results predicted on the basis of the
product ansatz are shown in Fig.~\ref{fig:s-+pi}.

%%%%%%%%%%%%%%%%%%%%%%%%%%%%%%%BEGIN-FIGURE%%%%
%%%%%%%%%%%%%%%%%%%%%%%%%%%%%%%%END-FIGURE%%%%%

The density of states has the same characteristic shape as seen previously.  The
spectral weight is strongly concentrated in the lowest finite-size excitation.
The scaling behavior of the transition rate for that state, $N|\langle
G|S_q^-|\lambda\rangle|^2\sim (1/N)^{1/ \eta-2}$, translates, via conformal invariance, into a
power-law infrared divergence,
\begin{equation}\label{eq:smp0}
S_{- +}^{\psi\psi}(\pi,\omega) \sim\omega^{1/ \eta-2}, \quad 1/ \eta-2=-1.346\ldots,
\end{equation}
for the spectral-weight distribution of the infinite chain. The solid line
represents a two-parameter fit, $a\omega^{1/ \eta-2}+b$, of the data points
representing the lowest excitation for $N=12, 16, 20, 24, 28$. The transition
rate data at higher frequencies appear to approach zero sufficiently rapidly to
overcome the divergent trend of the density of states to produce a monotonically
decreasing spectral-weight distribution with a cusp singularity at the upper
continuum boundary.

Similar single-peak lineshapes are expected for fixed-$q$ scans across the range
of the continuum P3. Hence the $\psi\psi$ contribution to the perpendicular spin
fluctuations is a structure that is strongly peaked along the lower continuum
boundary $\omega_L(q)$ in the shape of the psinon dispersion\cite{KM00} and a
shoulder reaching to the upper boundary $\omega_U(q)$ of the continuum P2.  Given
the strong divergence of $S_{- +}^{\psi\psi}(q,\omega)$ at $\omega_L(q)$, the
perpendicular spin fluctuations offer the most promising way to measure the
energy-momentum relation of the psinon quasiparticle by means of neutron
scattering.

%%%%%%%%%%%%%%%%%%%%%%%%%%%%%%%%%%%%%%%%%%%%%%%
%
\subsection{$S_{+ -}(q,\omega)$}\label{sec:P6}
%  
%%%%%%%%%%%%%%%%%%%%%%%%%%%%%%%%%%%%%%%%%%%%%%%

Here we are back to focusing on lineshapes produced by $\psi\psi^*$ excitations as
in Sec.~\ref{sec:P2}, but not in the same invariant $M_z$ subspace.
Nevertheless, the continuum P6 as depicted in Fig.~\ref{fig:pc6spec}(a)
produces, at $q=\pi/2$, a band of equal width and location as continuum P2
depicted in Fig.~\ref{fig:sc2} did.

\begin{figure}[tb]
 \centerline{\includegraphics[width=6.5cm,angle=-90]{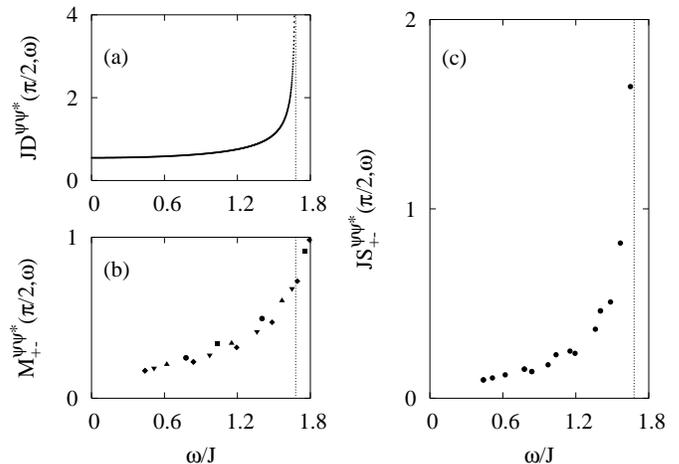}}

\caption{(a) Density of $\psi\psi$ states at $q=\pi/2$ for $N=2048$. (b)
  Perpendicular spin transition rates between the psinon vacuum and the $\psi\psi^*$
  states at $q=\pi/2$ for $N=12, 16, 20, 24, 28$. (c) Lineshape at $q=\pi/2$ of
  the $\psi\psi^*$ contribution to $S_{+ -}(q,\omega)$. All results are for $M_z=N/4$.}
\label{fig:s+-pid2}
\end{figure}

The data used in the product ansatz applied to $S_{+ -}^{\psi\psi^*}(\pi/2,\omega)$ are
shown in Fig.~\ref{fig:s+-pid2}. The density of states is exactly the same as in
Fig.~\ref{fig:sddpd2}. The data for the transition rates are monotonically
increasing with $\omega$. The trend in the low-frequency limit is that $M_{+
  -}^{\psi\psi^*}(\pi/2,\omega)$ approaches a finite value, possibly zero. Given the fact
that $D^{\psi\psi^*}(\pi/2,\omega)$ is flat at low frequencies, the same trend is observed
in $S_{+ -}^{\psi\psi^*}(\pi/2,\omega)$. The prediction of conformal invariance for the
leading infrared singularity is
\begin{equation}
S_{+ -}^{\psi\psi^*}(\pi/2,\omega) \sim \omega^{\eta+2},\quad \eta+2=3.531\ldots
\end{equation}

%%%%%%%%%%%%%%%%%%%%%%%%%%%%%%%BEGIN-FIGURE%%%%
%%%%%%%%%%%%%%%%%%%%%%%%%%%%%%%%END-FIGURE%%%%%

At the upper band edge $\omega_U$, the transition rate data exhibit a pronounced
maximum which could either signal a divergence or a cusp singularity for the
infinite system. Whatever the case may be, this enhancement amplifies the
divergent density of states in $S_{+ -}^{\psi\psi^*}(\pi/2,\omega)$.

Recall that the parallel spin fluctuations exhibit a double-peak lineshape at
the soft-mode wave number $q_s$. By contrast, the lineshape of the perpendicular
spin fluctuations at the soft-mode wave number $\bar{q}_s$ is a single-peak
structure with the spectral weight concentrated near the upper band edge.
For other wave numbers we do at present not have enough transitions rate data
points for a useful application of the product ansatz. Nevertheless, from the
few data points which we do have some interesting conclusions can be drawn.

In the transition rate data currently available we observe that the spectral
weight is heavily concentrated in a single excitation for any given $q\neq\pi/2$.
For $0\leq q<\bar{q}_s$ that excitation is located along the lower boundary of
continuum P6 and for $\bar{q}_s<q\leq\pi$ along the upper boundary. Both boundaries
have the shape of the antipsinon dispersion.\cite{KM00} Therefore, the
contribution of $S_{+ -}^{\psi\psi^*}(q,\omega)$ to the perpendicular spin fluctuations
offers the most direct way to measure the dispersion of the antipsinon
quasiparticle by means of neutron scattering -- not once but twice, in different
parts of the Brillouin zone.

\begin{figure}[htb]
\centering
\includegraphics[width=6.0cm,angle=-90]{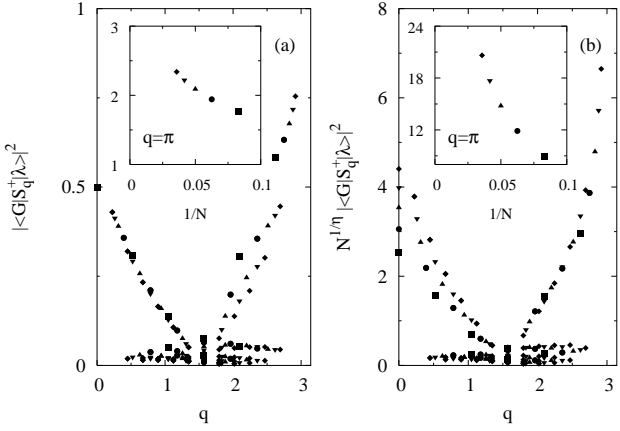}
\caption{(a) Unscaled and (b) scaled $\psi\psi^*$ transition rates for
  $S_{+ -}(q,\omega)$ at $M_z=N/4$ for $N=12, 16, 20, 24, 28$.}
\label{fig:mpc6+-}
\end{figure}

The outstanding role of the excitations along the two antipsinon branches at
$q<\bar{q}_s$ and $q>\bar{q}_s$ is illustrated in Fig.~\ref{fig:mpc6+-}, where
we have plotted the transition rates $|\langle G|S_q^+|\lambda\rangle|^2$ versus $q$ of all
states from continuum P6 across various system sizes. However, the role of the
finite-$N$ excitations which are part of these dominant branches is different at
$q<\bar{q}_s$ and $q>\bar{q}_s$. This is evident by comparison of panels (a) and
(b) which show differently scaled transition rates $N^\alpha|\langle G|S_q^+|\lambda\rangle|^2$.
    
%%%%%%%%%%%%%%%%%%%%%%%%%%%%%%%BEGIN-FIGURE%%%%
%%%%%%%%%%%%%%%%%%%%%%%%%%%%%%%%END-FIGURE%%%%%

To make sense of the data in this representation we must distinguish three
scenarios and ask the question: For what value of the scaling exponent $\alpha$ do
the transition rate data exhibit minimal $N$-dependence?

(i) For states that are inside a continuum, the product ansatz requires requires
that the exponent $\alpha=1$ minimizes the $N$-dependence of the transition rate
data. 

(ii) For states that mark the boundary of a continuum where the spectral-weight
distribution has a divergent singularity that is not caused by a divergent
density of states the exponent which minimizes the $N$-dependence of the
transition rate data is in the range $0<\alpha<1$.

(iii) For states that form a branch with nonvanishing spectral weight in the
limit $N\to\infty$, the exponent that minimizes the $N$-dependence is $\alpha=0$.

There is ample evidence for the first scenario in the results presented
earlier. The data in panel (a) strongly indicate that the third scenario is
realized for the antipsinon branch at $0<q<\bar{q}_s$. The evidence is rigorous
for the isolated excitation at $q=0$ which carries all the spectral of $S_{+
  -}(q,\omega)$ as discussed previously.

Panel (b) suggests that the second scenario applies to the antipsinon branch at
$\bar{q}_s<q\leq\pi$. The exponent $\alpha=1/ \eta=0.653\ldots$ used here is suggested by
conformal invariance, but there is a strong possibility that the singularity of
$S_{+ -}^{\psi\psi^*}(q,\omega)$ along the antipsinon branch at $\bar{q}_s<q\leq\pi$ is
governed by a $q$-dependent exponent. The insets to Fig.~\ref{fig:mpc6+-} show
the $N$-dependence of the corresponding data at $q=\pi$, which is off the scale
used in the main plots. At the zone boundary, the width of continuum P6 has
shrunk to zero, which is likely to affect the $N$-dependence of the transition
rate singularly, as indicated.

In conclusion, this study makes predictions of unprecedented detail for the
lineshapes of the spin and dimer fluctuations in the 1D $s=\frac{1}{2}$
Heisenberg antiferromagnet at zero temperature and nonzero magnetic
field. Foremost among these predictions is the direct observability in the
perpendicular spin fluctuations of the dispersion relations for the two
quasiparticles which play a crucial role in this situation: the psinon and the
antipsinon. 

%%%%%%%%%%%%%%%%%%%%%%%%%%%%%%%%%%%%%%%%%%%%%%%
%
\acknowledgments
%
%%%%%%%%%%%%%%%%%%%%%%%%%%%%%%%%%%%%%%%%%%%%%%%
Financial support  from the DFG Schwerpunkt 
\textit{Kollektive Quantenzust{\"a}nde in elektronischen 1D
{\"U}bergangsmetallverbindungen} (for M.K.) is gratefully acknowledged.

\begin{appendix}
%%%%%%%%%%%%%%%%%%%%%%%%%%%%%%%%%%%%%%%%%%%%%
%
\section{Calculating matrix elements via Bethe ansatz}\label{appA}
%
%%%%%%%%%%%%%%%%%%%%%%%%%%%%%%%%%%%%%%%%%%%%%
The Bethe ansatz\cite{Beth31} is an exact method for the calculation of
eigenvectors of integrable quantum many-body systems. The Bethe wave function of
any eigenstate of \eqref{eq:Hh} in the invariant subspace with $r=N/2-M_z$
reversed spins relative to the magnon vacuum,
\begin{equation}\label{eq:psir}
 |\psi\rangle = \sum_{1\leq n_1<\ldots<n_r\leq N} a(n_1,\ldots,n_r)
 S_{n_1}^-\cdots S_{n_r}^-|F\rangle,
\end{equation}
has coefficients of the form
\begin{equation}\label{eq:bar}
 a(n_1,\ldots,n_r) =
 \!\sum_{{\cal P}\in S_r}
 \!\exp \left(\!i\sum_{j=1}^r k_{{\cal P} j}n_j
 + \frac{i}{2}\sum_{i<j}^{r} \theta_{{\cal P}i{\cal P}j}\!\right)
\end{equation}
determined by $r$ magnon momenta $k_i$ and one phase angle
$\theta_{ij}=-\theta_{ji}$ for each magnon pair. The sum ${\cal P}\in S_r$ is
over the permutations of the labels $\{1,2,\ldots,r\}$. 

The consistency requirements for the coefficients $a(n_1,\ldots,n_r)$ inferred from
the eigenvalue equation $H|\psi\rangle=E|\psi\rangle$ and the requirements imposed
by translational invariance lead to a set of coupled nonlinear
equations for the $k_i$ and $\theta_{ij}$. A computationally
convenient rendition of the Bethe ansatz equations for a state specified by
Bethe quantum numbers $I_1,\ldots,I_r$ has the form 
\begin{equation}
\label{eq:bae}
N\phi(z_i) = 2\pi I_i + \sum_{j\neq i}\phi\bigl [(z_i-z_j)/2\bigr ],
\quad i=1,\ldots,r,
\end{equation}
where $\phi(z) \equiv 2\arctan z$, $k_i = \pi -\phi(z_i)$ and $\theta_{ij} = \pi \, {\rm
  sgn}[{\Re}(z_i-z_j)] - \phi\bigl[(z_i-z_j)/2\bigr]$. The energy and wave number
of the eigenvector thus determined are
\begin{equation}\label{eq:ekz}
\frac{E-E_F}{J} = -\sum_{i=1}^{r}\frac{2}{1+z_i^2},\quad
k = \pi r - \frac{2\pi}{N}\sum_{i=1}^r I_i,
\end{equation}
where $E_F=JN/4$ is the energy of the magnon vacuum.

In the past, The Bethe ansatz was rarely used for the purpose of calculating
matrix elements. The main deterrent has been the need of evaluating the sum ${\cal
  P}\in S_r$ over the $r!$ magnon permutations in the coefficients \eqref{eq:bar} of the
Bethe eigenvectors \eqref{eq:psir}. However, the tide is now changing rapidly
for two reasons: (i) the availability of vastly higher computational power, (ii)
theoretical advances that make it possible to reduce matrix elements of Bethe
wave functions to determinantal expressions.\cite{KMT99}

In the following, we sketch how the matrix elements can be manipulated
effectively by using the Bethe wave functions directly. The use of the
determinantal expressions of Bethe ansatz transition rates for the calculation
of dynamic structure factors will be reported elsewhere.\cite{KBM02u}

In designing an efficient algorithm, we must heed the fact that in the
calculation of a single matrix element, the sum ${\cal P}\in S_r$ is evaluated
many times, once for every coefficient $a(n_1,\ldots,n_r)$ of the two
eigenvectors involved. Under these circumstances, it is imperative that the
algorithm has rapid access to a table of permutations. Such tables can be
generated recursively by powerful algorithms.\cite{Sedg92}

The computational effort can be reduced considerably if we use the translational
symmetry of \eqref{eq:psir}, guaranteed by the relation $a(n_1+l,\ldots,n_r+l) =
e^{ikl} a(n_1,\ldots,n_r)$ between Bethe coefficients \eqref{eq:bar} pertaining to
basis vectors that transform into each other under translation. Translationally
invariant basis vectors have the form
\begin{equation}\label{eq:tibv}
|j;k\rangle \equiv \frac{1}{\sqrt{d_j}}\sum_{l=0}^{d_j-1}e^{ilk}|j\rangle_l,
\end{equation}
where $|j\rangle_l \equiv {\bf T}^l|j\rangle_0 =
|n_1^{(j)}-l,\ldots,n_r^{(j)}-l\rangle$  and $1\leq N/d_j\leq
N$ is an integer. The wave numbers $k$ realized in the set (\ref{eq:tibv}) are
multiples mod($2\pi$) of $2\pi/d_j$.

The set of basis vectors $|j\rangle_0 = |n_1^{(j)},\ldots,n_r^{(j)}\rangle, j=1,\ldots,d$, are the
generators of the translationally invariant basis. The set of distinct vectors
$|j;k\rangle$ for fixed $k$ is labeled $j\in{\cal J}_k \subseteq \{1,\ldots,d\}$. The rotationally
invariant subspace for fixed $N/2-r$, which has dimensionality $D =
N!/[r!(N-r)!]$, splits into $N$ translationally invariant subspaces of
dimensionality $D_k$, one for each wave number $k=2\pi n/N, n=0,\ldots,N-1$. We have
\begin{equation}\label{eq:djDk}
D = \sum_{j=1}^{d}d_j = \sum_{0\leq k< 2\pi} D_k,~~ D_k = \sum_{j\in{\cal J}_k}.
\end{equation}

The Bethe eigenvector (\ref{eq:psir}) expanded in this basis can thus be
written in the form
\begin{equation}\label{eq:psik}
|\psi\rangle = \sum_{j\in{\cal J}_k}a_j\sum_{l=0}^{d_j-1}e^{ilk}|j\rangle_l
= \sum_{j=1}^da_j\sum_{l=0}^{d_j-1}e^{ilk}|j\rangle_l, 
\end{equation}
where the $a_j \equiv a(n_1^{(j)},\ldots,n_r^{(j)})$, the Bethe coefficients of the
generator basis vectors $|j\rangle_0$, are the only ones that must be evaluated. The
last expression of (\ref{eq:psik}) holds because the Bethe coefficients $a_j$ of
all generators $|j\rangle_0$ which do not occur in the set ${\cal J}_k$ are zero.

We calculate transition rates for the dynamic structure factor (\ref{eq:dssf})
in the form
\begin{equation}\label{eq:mes}
|\langle G|S_q^\mu|m\rangle|^2 = \frac{|\langle\psi_0|S_q^\mu|\psi_m\rangle|^2} {||\psi_0||^2 ||\psi_m||^2},
\end{equation}
where $|\psi_0\rangle, |\psi_m\rangle$ are the (non-normalized) Bethe eigenvectors of the
ground state and of one of the excited states from classes (i)-(vi),
respectively.  The norms are
\begin{equation}\label{eq:ba-norm}
|| \psi ||^{2} = \sum_{j=1}^d d_j |a_j|^{2}.
\end{equation}
The matrix element $\langle\psi_0|S_q^\mu|\psi_m\rangle$ is nonzero only if
$q=k_{m}-k_{0}+2\pi\mathbb{Z}$. For the fluctuation operator $S_q^z$ it can be
evaluated in the form:
\begin{eqnarray}\label{eq:matrix-sq}
  && \langle \psi_0 |S_q^{z}|\psi_m \rangle \!=\! 
  \frac{1}{\sqrt{N}}\sum_{j=1}^{d}\bar{a}_{j}^{(0)}a_{j}^{(m)}\sum_{n=1}^{N} 
  e^{iqn}\sum_{l=0}^{d_j-1} e^{ilq} {_l}\langle j | S_n^{z} | j\rangle_{l}.
    \nonumber \\ 
\end{eqnarray}
The non-vanishing matrix elements $\langle\psi_0|S_q^\pm|\psi_m\rangle$ for the other spin
fluctuation operators must also satisfy $q=k_{m}-k_{0}+2\pi\mathbb{Z}$ and can be
reduced to somewhat more complicated expressions involving elements $_{l_0}\langle
j_0|S_n^\pm|j_m\rangle_{l_m}$ between basis vectors from different $S_T^z$ subspaces.

The memory requirements for the calculation of one such matrix element are
6.7MB for $N=18,r=9$ and 73MB for $N=20,r=10$.

\end{appendix}

%%%%%%%%%%%%%%%%%%%%%%%%%%%%%%%%%%%%%%%%%%%%%%%
%\ifthenelse{\equal{\writer}{gerhard}}%
%{\bibliography{../references,notes}}%
%{\bibliography{/home/karbach/REFERENCES/references,notes}}

\begin{thebibliography}{100}
\expandafter\ifx\csname bibnamefont\endcsname\relax
  \def\bibnamefont#1{#1}\fi
\expandafter\ifx\csname bibfnamefont\endcsname\relax
  \def\bibfnamefont#1{#1}\fi
\expandafter\ifx\csname url\endcsname\relax
  \def\url#1{\texttt{#1}}\fi
\expandafter\ifx\csname urlprefix\endcsname\relax\def\urlprefix{URL }\fi
\providecommand{\bibinfo}[2]{#2}
\providecommand{\eprint}[2][]{\url{#2}}

\bibitem{MK00}
\bibinfo{author}{\bibfnamefont{G.}~\bibnamefont{M{\"u}ller}} \bibnamefont{and}
  \bibinfo{author}{\bibfnamefont{M.}~\bibnamefont{Karbach}}, in
  \emph{\bibinfo{booktitle}{Frontiers of Neutron Scattering}}, edited by
  \bibinfo{editor}{\bibfnamefont{A.}~\bibnamefont{Furrer}}
  (\bibinfo{publisher}{World Scientific}, \bibinfo{address}{Singapore},
  \bibinfo{year}{2000}), p. \bibinfo{pages}{168}, \eprint{cond-mat/0003076}.

\bibitem{KBI93}
\bibinfo{author}{\bibfnamefont{V.~E.} \bibnamefont{Korepin}},
  \bibinfo{author}{\bibfnamefont{N.~M.} \bibnamefont{Bogoliubov}},
  \bibnamefont{and} \bibinfo{author}{\bibfnamefont{A.~G.}
  \bibnamefont{Izergin}}, \emph{\bibinfo{title}{Quantum Inverse Scattering
  Method and Correlation Functions}} (\bibinfo{publisher}{Cambridge University
  Press}, \bibinfo{address}{Cambridge}, \bibinfo{year}{1993}).

\bibitem{GRS96}
\bibinfo{author}{\bibfnamefont{C.}~\bibnamefont{G{\'o}mez}},
  \bibinfo{author}{\bibfnamefont{M.}~\bibnamefont{Ruiz-Altaba}},
  \bibnamefont{and} \bibinfo{author}{\bibfnamefont{G.}~\bibnamefont{Sierra}},
  \emph{\bibinfo{title}{Quantum Groups in Two-Dimensional Physics}}
  (\bibinfo{publisher}{Cambridge University Press},
  \bibinfo{address}{Cambridge}, \bibinfo{year}{1966}).

\bibitem{Beth31}
\bibinfo{author}{\bibfnamefont{H.}~\bibnamefont{Bethe}}, \bibinfo{journal}{Z.
  Phys.} \textbf{\bibinfo{volume}{71}}, \bibinfo{pages}{205}
  (\bibinfo{year}{1931}).

\bibitem{FT81}
\bibinfo{author}{\bibfnamefont{L.~D.} \bibnamefont{Faddeev}} \bibnamefont{and}
  \bibinfo{author}{\bibfnamefont{L.~A.} \bibnamefont{Takhtajan}},
  \bibinfo{journal}{Phys. Lett.} \textbf{\bibinfo{volume}{A85}},
  \bibinfo{pages}{375} (\bibinfo{year}{1981}).

\bibitem{KM00}
\bibinfo{author}{\bibfnamefont{M.}~\bibnamefont{Karbach}} \bibnamefont{and}
  \bibinfo{author}{\bibfnamefont{G.}~\bibnamefont{M{\"u}ller}},
  \bibinfo{journal}{Phys. Rev. B} \textbf{\bibinfo{volume}{62}},
  \bibinfo{pages}{14871} (\bibinfo{year}{2000}).

\bibitem{HSR+99}
\bibinfo{author}{\bibfnamefont{P.~R.} \bibnamefont{Hammar}},
  \bibinfo{author}{\bibfnamefont{M.~B.} \bibnamefont{Stone}},
  \bibinfo{author}{\bibfnamefont{D.~H.} \bibnamefont{Reich}},
  \bibinfo{author}{\bibfnamefont{C.}~\bibnamefont{Broholm}},
  \bibinfo{author}{\bibfnamefont{P.~J.} \bibnamefont{Gibson}},
  \bibinfo{author}{\bibfnamefont{M.~M.}~\bibnamefont{Turnbull}},
  \bibinfo{author}{\bibfnamefont{C.~P.} \bibnamefont{Landee}},
  \bibnamefont{and} \bibinfo{author}{\bibfnamefont{M.}~\bibnamefont{Oshikawa}},
  \bibinfo{journal}{Phys. Rev. B} \textbf{\bibinfo{volume}{59}},
  \bibinfo{pages}{1008} (\bibinfo{year}{1999}).

\bibitem{AFM+96}
\bibinfo{author}{\bibfnamefont{M.}~\bibnamefont{Arai}},
  \bibinfo{author}{\bibfnamefont{M.}~\bibnamefont{Fujita}},
  \bibinfo{author}{\bibfnamefont{M.}~\bibnamefont{Motokawa}},
  \bibinfo{author}{\bibfnamefont{J.}~\bibnamefont{Akimitsu}}, \bibnamefont{and}
  \bibinfo{author}{\bibfnamefont{S.~M.} \bibnamefont{Bennington}},
  \bibinfo{journal}{Phys. Rev. Lett.} \textbf{\bibinfo{volume}{77}},
  \bibinfo{pages}{3649} (\bibinfo{year}{1996}).

\bibitem{FKL+98}
\bibinfo{author}{\bibfnamefont{K.}~\bibnamefont{Fabricius}},
  \bibinfo{author}{\bibfnamefont{A.}~\bibnamefont{Kl{\"u}mper}},
  \bibinfo{author}{\bibfnamefont{U.}~\bibnamefont{L{\"o}w}},
  \bibinfo{author}{\bibfnamefont{B.}~\bibnamefont{B{\"u}chner}},
  \bibinfo{author}{\bibfnamefont{T.}~\bibnamefont{Lorenz}},
  \bibinfo{author}{\bibfnamefont{G.}~\bibnamefont{Dhaleene}}, \bibnamefont{and}
  \bibinfo{author}{\bibfnamefont{A.}~\bibnamefont{Revcolevschi}},
  \bibinfo{journal}{Phys. Rev. B} \textbf{\bibinfo{volume}{57}},
  \bibinfo{pages}{1102} (\bibinfo{year}{1998}).

\bibitem{FL98}
\bibinfo{author}{\bibfnamefont{K.}~\bibnamefont{Fabricius}} \bibnamefont{and}
  \bibinfo{author}{\bibfnamefont{U.}~\bibnamefont{L{\"o}w}},
  \bibinfo{journal}{Phys. Rev. B} \textbf{\bibinfo{volume}{57}},
  \bibinfo{pages}{13371} (\bibinfo{year}{1998}).

\bibitem{note1}
\bibinfo{note}{The approximate nature of these additional selection rules
  reflects a nontrivial relationship between psinon creation operators and the
  physically motivated fluctuation operators. In the Haldane-Shastry model
  (Ref.~\onlinecite{TH94}), the relationship between quasiparticle creation
  operator and spin fluctuation operator is more direct. Here some of the
  additional selection rules are rigorous.}

\bibitem{MTBB81}
\bibinfo{author}{\bibfnamefont{G.}~\bibnamefont{M{\"u}ller}},
  \bibinfo{author}{\bibfnamefont{H.}~\bibnamefont{Thomas}},
  \bibinfo{author}{\bibfnamefont{H.}~\bibnamefont{Beck}}, \bibnamefont{and}
  \bibinfo{author}{\bibfnamefont{J.~C.} \bibnamefont{Bonner}},
  \bibinfo{journal}{Phys. Rev. B} \textbf{\bibinfo{volume}{24}},
  \bibinfo{pages}{1429} (\bibinfo{year}{1981}).

\bibitem{note2}
\bibinfo{note}{There is one exception to that rule. The sole excitation that
  contributes to $S_{+-}(0,\omega)$ is a class (v) excitation at $\omega=h$
  (Ref.~\onlinecite{MTBB81}).}

\bibitem{Hald80}
\bibinfo{author}{\bibfnamefont{F.~D.~M.} \bibnamefont{Haldane}},
  \bibinfo{journal}{Phys. Rev. Lett.} \textbf{\bibinfo{volume}{45}},
  \bibinfo{pages}{1358} (\bibinfo{year}{1980}).

\bibitem{FGM+96}
\bibinfo{author}{\bibfnamefont{A.}~\bibnamefont{Fledderjohann}},
  \bibinfo{author}{\bibfnamefont{C.}~\bibnamefont{Gerhardt}},
  \bibinfo{author}{\bibfnamefont{K.-H.} \bibnamefont{M{\"u}tter}},
  \bibinfo{author}{\bibfnamefont{A.}~\bibnamefont{Schmitt}}, \bibnamefont{and}
  \bibinfo{author}{\bibfnamefont{M.}~\bibnamefont{Karbach}},
  \bibinfo{journal}{Phys. Rev. B} \textbf{\bibinfo{volume}{54}},
  \bibinfo{pages}{7168} (\bibinfo{year}{1996}).

\bibitem{KMB+97}
\bibinfo{author}{\bibfnamefont{M.}~\bibnamefont{Karbach}},
  \bibinfo{author}{\bibfnamefont{G.}~\bibnamefont{M{\"u}ller}},
  \bibinfo{author}{\bibfnamefont{A.~H.} \bibnamefont{Bougourzi}},
  \bibinfo{author}{\bibfnamefont{A.}~\bibnamefont{Fledderjohann}},
  \bibnamefont{and} \bibinfo{author}{\bibfnamefont{K.-H.}
  \bibnamefont{M{\"u}tter}}, \bibinfo{journal}{Phys. Rev. B} \textbf{\bibinfo{volume}{55}},
  \bibinfo{pages}{12510} (\bibinfo{year}{1997}).

\bibitem{BKM98}
\bibinfo{author}{\bibfnamefont{A.~H.} \bibnamefont{Bougourzi}},
  \bibinfo{author}{\bibfnamefont{M.}~\bibnamefont{Karbach}}, \bibnamefont{and}
  \bibinfo{author}{\bibfnamefont{G.}~\bibnamefont{M{\"u}ller}},
  \bibinfo{journal}{Phys. Rev. B} \textbf{\bibinfo{volume}{57}},
  \bibinfo{pages}{11429} (\bibinfo{year}{1998}).

\bibitem{KHM98}
\bibinfo{author}{\bibfnamefont{M.}~\bibnamefont{Karbach}},
  \bibinfo{author}{\bibfnamefont{K.}~\bibnamefont{Hu}}, \bibnamefont{and}
  \bibinfo{author}{\bibfnamefont{G.}~\bibnamefont{M{\"u}ller}},
  \bibinfo{journal}{Comp. in Phys.} \textbf{\bibinfo{volume}{12}},
  \bibinfo{pages}{565} (\bibinfo{year}{1998}).

\bibitem{KMT99}
\bibinfo{author}{\bibfnamefont{N.}~\bibnamefont{Kitanine}},
  \bibinfo{author}{\bibfnamefont{J.~M.} \bibnamefont{Maillet}},
  \bibnamefont{and} \bibinfo{author}{\bibfnamefont{V.}~\bibnamefont{Terras}},
  \bibinfo{journal}{Nucl. Phys. B} \textbf{\bibinfo{volume}{554}},
  \bibinfo{pages}{647} (\bibinfo{year}{1999}).

\bibitem{KBM02u}
\bibinfo{note}{M. Karbach, D. Biegel, and G. M{\"u}ller (unpublished).}

\bibitem{Sedg92}
\bibinfo{author}{\bibfnamefont{R.}~\bibnamefont{Sedgewick}},
  \emph{\bibinfo{title}{Algorithms in C++}} (\bibinfo{publisher}{Addison
  Wesley}, \bibinfo{address}{Reading, Massachusetts}, \bibinfo{year}{1992}).

\bibitem{TH94}
\bibinfo{author}{\bibfnamefont{J.~C.} \bibnamefont{Talstra}} \bibnamefont{and}
  \bibinfo{author}{\bibfnamefont{F.~D.~M.} \bibnamefont{Haldane}},
  \bibinfo{journal}{Phys. Rev. B} \textbf{\bibinfo{volume}{50}},
  \bibinfo{pages}{6889} (\bibinfo{year}{1994}).

\end{thebibliography}
%\bibliographystyle{apsrev}

%%%%%%%%%%%%%%%%%%%%%%%%%%%%%%%%%%%%%%%%%%%%%%%
\end{document}